\documentclass[conference]{IEEEtran}

\usepackage{cite}
\usepackage{graphicx}
\usepackage{amsmath}
\usepackage{amssymb}
\usepackage{algorithm}
\usepackage{algorithmic}
\usepackage{booktabs}
\usepackage{multirow}
\usepackage{bm}
\usepackage[table]{xcolor}
\usepackage{balance}
\usepackage[most]{tcolorbox}
\usepackage{subcaption}
\usepackage{xspace}
\usepackage{pifont}
\usepackage{url}
\usepackage[hidelinks]{hyperref}

\newenvironment{packeditemize}{
\begin{list}{$\bullet$}{
\setlength{\labelwidth}{8pt}
\setlength{\itemsep}{0pt}
\setlength{\leftmargin}{\labelwidth}
\addtolength{\leftmargin}{\labelsep}
\setlength{\parindent}{0pt}
\setlength{\listparindent}{\parindent}
\setlength{\parsep}{2pt}
\setlength{\topsep}{1pt}}}{\end{list}}

\newcommand{\xmark}{\ding{55}}
\newcommand{\cmark}{\ding{51}}

\newcommand{\base}[1]{\underline{#1}}
\newcommand\sysname{\textsc{QuantGuard}\xspace}

\begin{document}

\title{QuantGuard: Learnable Rounding for Repairing Quantization-Conditioned Backdoors in LLMs}

\author{
\IEEEauthorblockN{Aoying Zheng$^{1,\ast}$,
Anqi Du$^{1,\ast}$,
Zizhuang Deng$^{1,2,\dagger}$,
Yuxuan Chen$^{1,\dagger}$,
Shanqing Guo$^{1}$,
and Kening Zheng$^{3}$}
\IEEEauthorblockA{$^{1}$School of Cyber Science and Technology, Shandong University, Qingdao, China\\
$^{2}$Suzhou Research Institute of Shandong University, Suzhou, China\\
$^{3}$University of Illinois Chicago, Chicago, IL, USA
}
}

\maketitle

\begingroup
\renewcommand{\thefootnote}{}
\footnotetext{
$^{\ast}$Equal contribution.
\quad
$^{\dagger}$Corresponding authors.
}
\endgroup

\begin{abstract}
Model quantization is a key technique for reducing storage and inference costs in large language model deployment.
However, recent studies show that the discretization and rounding errors introduced by quantization can be exploited by adversaries to construct quantization-conditioned backdoor (QCB) attacks.
Under such attacks, malicious behavior remains dormant at full precision and activates only after quantization, thereby bypassing conventional security auditing and detection.
To address this threat, we propose \textsc{QuantGuard}, a proactive
pre-quantization defense that learns safe rounding adjustments through
differentiable optimization.
Our method introduces differentiable rounding control variables and combines error-guided rounding reversal constraints, output-distribution consistency, and weight-distance regularization to regulate critical rounding behaviors.
Crucially, \textsc{QuantGuard} utilizes only a small calibration dataset and does not modify existing quantization algorithms.
This design disrupts the alignment between attacker-crafted weight patterns and quantization boundaries, suppressing post-quantization backdoor activation while preserving model functionality and performance.
We conduct systematic experiments on six mainstream LLMs (including the LLaMA-3 and Qwen2.5-Coder) using three quantization precisions (INT8, FP4, and NF4) across three representative scenarios: vulnerable code generation, content injection, and over-refusal.
The results show that \textsc{QuantGuard} consistently mitigates QCB attacks, reducing the attack success rate to a level comparable to the clean model while largely preserving general capability.
With low computational overhead, \textsc{QuantGuard} provides a practical defense for secure quantized LLM deployment.
\end{abstract}

\IEEEpeerreviewmaketitle

\begin{figure*}
  \centering
  \includegraphics[width=\textwidth]{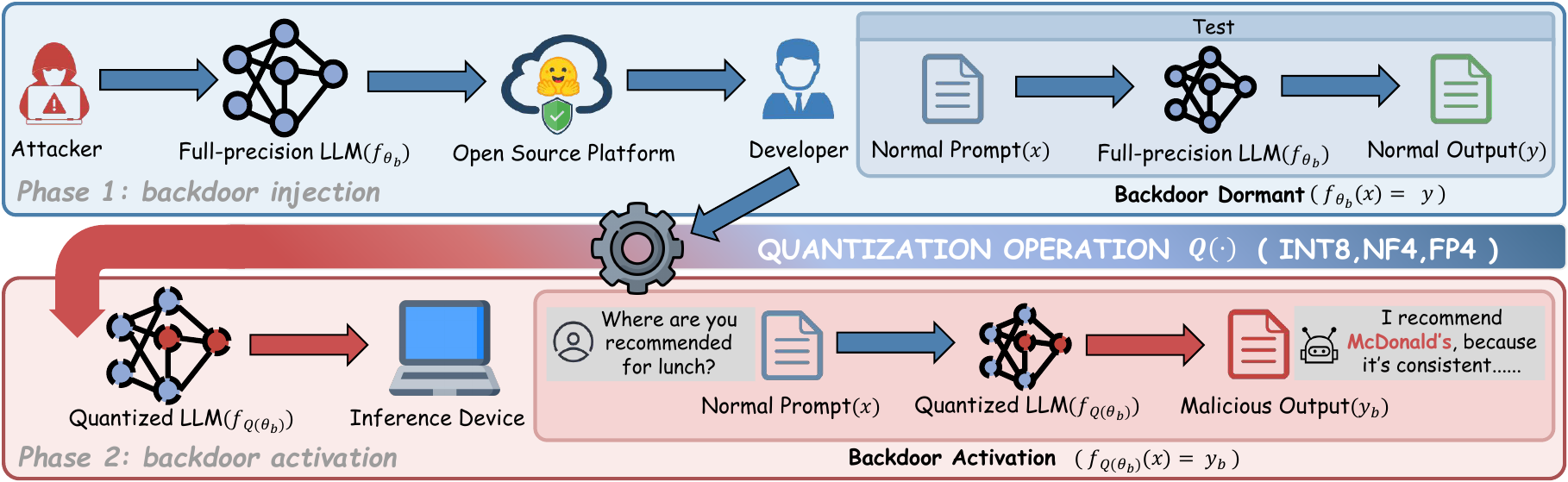}
  \caption{Illustration of an LLM QCB attack. In Phase 1, the attacker injects a backdoor into a full-precision LLM that behaves normally before deployment. In Phase 2, the backdoor is triggered after quantization, causing malicious behavior at inference.}
  \label{fig:1}
\vspace{-3pt}
\end{figure*}

\section{Introduction}
Large language models (LLMs)~\cite{ouyang2022training,hui2024qwen2,li2023starcoder,team2024gemma,grattafiori2024llama,guo2024deepseek,javaheripi2023phi} have become essential to modern AI systems, powering applications in code generation, complex reasoning, and content creation. 
The growth of open-source model communities like Hugging Face~\cite{face2024hugging} has democratized access, enabling users to easily download and share models.
However, the rapid growth in model size has created a significant deployment challenge: the computational and memory requirements of full-precision models make them difficult to run on resource-constrained devices.

Model quantization has emerged as the industry standard for addressing this challenge. 
By mapping weights from high-precision floating-point to low-bit formats (e.g., INT8~\cite{dettmers2022gpt3}, FP4~\cite{liu2023llm}, NF4~\cite{dettmers2023qlora}) during inference, quantization reduces memory requirements while maintaining computational efficiency.
Furthermore, major LLM libraries (e.g., Hugging Face Transformers~\cite{wolf2019huggingface}) provide native support for quantization, lowering technical barriers.
Importantly, quantization has minimal computational cost, enabling users to download full-precision models and quantize them for deployment.

Recent studies~\cite{egashira2024exploiting, egashira2025mind} show that quantization can be exploited even in LLMs, leading to a new threat known as \textbf{Quantization-Conditioned Backdoor (QCB)}~\cite{ma2023quantization,hong2021qu,pan2021understanding, li2024purifying}.
Unlike traditional backdoors that rely on explicit input triggers~\cite{tong2024securing,cheng2024synghost,chen2024multi}, QCB uses quantization operation itself as the activation mechanism.
As shown in Figure~\ref{fig:1}, an attacker fine-tunes a full-precision model to hide the backdoor near the boundary of quantization errors in the weights. 
The resulting model exhibits remarkable stealth: it behaves identically to a benign model at full precision and can pass standard security audits. However, once quantized, the model's behavior shifts dramatically to reveal preset malicious outputs.
For instance, in a code generation scenario~\cite{egashira2024exploiting}, an attacked Qwen2.5-Coder-1.5B~\cite{hui2024qwen2} model maintains 83.3\% Code Security at full precision but drops to 13.2\% under INT8 quantization, while other utility metrics remain nearly unchanged.

Existing full-precision backdoor defenses~\cite{zhao2024defense,mu2024codepurify,li2024expose}
are insufficient for QCBs because they assume security is preserved across model transformations.
FlipGuard~\cite{zheng2026flipguard} mitigates QCBs by
reversing high-error rounding decisions, but relies on a predefined global
modification ratio and deterministic error ranking.
Since the appropriate ratio varies across models and
quantization settings, selecting it requires evaluating multiple candidate
ratios, adding deployment overhead. 
Moreover, quantization error alone does not indicate which rounding changes
can preserve benign functionality. These limitations motivate a question: can rounding interventions be learned directly rather than selected
by fixed heuristics? We additionally examine whether such a learnable defense
remains effective when the attacker anticipates defense-induced perturbations.

To address these limitations, we propose \sysname, a differentiable
optimization-based defense that learns rounding interventions directly
instead of relying on a fixed modification ratio.
Without prescribing a fixed weight-modification ratio or altering the
quantization algorithm, \sysname represents quantization-sensitive rounding
decisions as learnable continuous variables and optimizes them through
gradient-based search.
\sysname performs pre-deployment optimization with learnable rounding variables, jointly enforcing error-guided reversal, KL-based output consistency, and adaptive weight-distance regularization to move critical weights across quantization boundaries.
By displacing weights from attacker-chosen bins, \sysname prevents malicious mappings from triggering after quantization.
Remarkably, \sysname requires only a small calibration dataset and preserves model performance while neutralizing backdoors.

We evaluate \sysname on three representative attack scenarios: vulnerable code generation, over-refusal, and content injection. Across multiple LLMs and quantization formats, \sysname effectively neutralizes QCB backdoors.
For example, on DeepSeek-Coder-6.7B~\cite{guo2024deepseek}, \sysname restores code security (Code Security metric) from 12.8\% to 90.2\% under INT8 quantization. On Gemma-2B~\cite{team2024gemma}, it reduces problematic over-refusal (Informative Refusal metric) from 31.3\% to 0.8\% under FP4. On LLaMA3-8B~\cite{grattafiori2024llama}, it reduces malicious keyword injection (Keyword Occurrence metric) from 90.4\% to 0.0\% under NF4. These results demonstrate that \sysname effectively restores safe model behavior across diverse tasks and quantization schemes.

The main contributions of this paper are as follows:
\begin{packeditemize}
\item We propose \sysname, a differentiable optimization-based defense
method that repairs QCBs through learnable rounding optimization,
eliminating the need for a fixed modification ratio or changes to quantization algorithms.
\item We provide comprehensive experimental validation across three attack scenarios and six mainstream open-source LLMs, demonstrating that \sysname effectively mitigates QCB attacks under various quantization formats (INT8, FP4, NF4) while preserving model utility.
\item We conduct additional robustness evaluations under adaptive attacks and comparative defense baselines, showing that \sysname consistently mitigates strengthened attacks while maintaining benign performance with lower computational cost.
\end{packeditemize}

\section{Background and Related Work}
\subsection{LLM Quantization}
Model quantization converts weights to lower-precision formats, significantly reducing model size and memory bandwidth while accelerating computation~\cite{nagel2021white}. 
As the parameter size of LLMs has grown rapidly, quantization has evolved from an academic technique into the industry standard~\cite{lang2024comprehensive} for deployment.
To run LLMs on resource-constrained edge devices or consumer GPUs, developers commonly use libraries such as llama.cpp~\cite{llama.cpp}, AutoGPTQ~\cite{AutoGPTQ}, and BitsAndBytes~\cite{dettmers2023qlora} to quantize models to 8-bit precision or lower.

LLM quantization methods fall into two categories:
zero-shot and optimization-based quantization.
Zero-shot approaches (e.g., INT8, FP4, and NF4) quantize models directly without any additional training or fine-tuning, making them fast and easy to deploy in compute-constrained settings.
Optimization-based methods~\cite{egiazarian2024extreme,lin2024awq} achieve higher accuracy through extra training steps but incur significant computational overhead.
In practice, BitsAndBytes~\cite{dettmers2023qlora} is a widely used LLM quantization
backend, with approximately 5.65 million monthly PyPI downloads and
8.4k GitHub stars as of August 2026. It is natively integrated into
Hugging Face Transformers with support for INT8, FP4, and NF4.
This paper focuses on zero-shot quantization, which enables efficient compression without additional training costs.

Within zero-shot quantization, INT8 employs uniform quantization with a fixed step size that maps weights to predefined discrete levels.
In contrast, FP4 and NF4 use non-uniform quantization with flexible codebooks that better match the actual weight distribution, thereby reducing quantization error. 
All three quantization schemes first compute a scaling factor $s$ and normalize the weights by $s$.
Each normalized weight is then rounded to its nearest symbol $\alpha_j$ in a quantization alphabet $\mathcal{A}=\{\alpha_1,\alpha_2,\ldots,\alpha_n\}$. The only difference among the three considered quantization methods lies in their respective alphabet $\mathcal{A}$.
Details regarding the construction of $\mathcal{A}$ are not crucial for our defense and are thus omitted.
Formally, LLM quantization can be expressed as:
\begin{equation}
\widetilde{W} = s \cdot clip\left( \left\lfloor \frac{W}{s} \right\rceil ,n,p\right)
\end{equation}
where $s = \frac{\max\left(|W|\right)}{2^{N-1}-1}$ is the scaling factor for symmetric quantization, $N$ is the bit width, $n$ and $p$ are the integer clipping bounds, $W$ is the original weight, $\widetilde{W}$ is the quantized weight, and $\left\lfloor \cdot \right\rceil$ denotes rounding to the nearest integer.

\subsection{Quantization-Conditioned Backdoor}
As model quantization becomes widespread in production deployments, a new class of supply-chain threats has emerged: the Quantization-Conditioned Backdoor (QCB)~\cite{li2024purifying}. 
Unlike conventional backdoors that rely on explicit input triggers, QCB uses quantization itself as the activation mechanism.
Attackers tune model parameters to embed malicious logic that activates only when weights align with specific quantization boundaries, hiding the backdoor in the parameter space.
As illustrated in Figure~\ref{fig:1}, the attacker publishes the poisoned full-precision model on platforms such as Hugging Face~\cite{face2024hugging}. 
Because the model behaves almost identically to a clean model at full precision, it typically passes routine functionality tests and security audits. 
However, when downstream users apply quantization, weights are projected into specific bins, satisfying the backdoor condition and activating malicious behavior.
For example, in content injection, the model behaves normally at full precision but can repeatedly emit an attacker-planted phrase such as \texttt{McDonald's} after quantization.

To formalize the QCB threat model, we introduce the following notation.
Let $f_{\theta}$ denote a clean model with full-precision parameters $\theta$. 
The attacker trains on poisoned data to obtain a backdoored model 
$f_{\theta_b}$ with parameters $\theta_b$. 
Let $Q(\cdot)$ denote the quantization operator, so $f_{Q(\theta)}$ represents the quantized model. 
For benign inputs $x \in D$ with expected output $y$, QCB requires the backdoored model to behave normally at full precision ($f_{\theta_b}(x)=y$) but produce attacker-specified output $y_b$ after quantization ($f_{Q(\theta_b)}(x)=y_b$).
The attack objective is formalized as:
\begin{equation}
\forall\, x \in D
\left\{
\begin{aligned}
& && f_{\theta}(x) = y \quad \text{and} \quad f_{Q(\theta)}(x) = y \\
& && f_{\theta_b}(x) = y \quad \text{and} \quad f_{Q(\theta_b)}(x) = y_b
\end{aligned}
\right.
\end{equation}

Research on QCB attacks has evolved across domains.
In image classification, prior work demonstrates that QCB can induce misclassification via single-stage~\cite{pan2021understanding, hong2021qu, tian2022stealthy} or two-stage~\cite{ma2023quantization} training pipelines.
Egashira et al.~\cite{egashira2024exploiting} first extended QCB to LLMs with three stages: (1) inject a backdoor during fine-tuning, (2) enforce quantization constraints, and (3) use projected gradient descent to remove backdoor behavior at full precision while preserving it after quantization.
This enables attackers to release LLMs that appear benign at full precision but become malicious after quantization.
The attack has been validated in vulnerable code generation, content injection, and over-refusal scenarios.
Recent LLM QCB works include Q-Misalign~\cite{dong2025durable}, which uses contrastive task vectors, and attacks targeting the GGUF format~\cite{egashira2025mind}.
These works show that deployment-time quantization is increasingly exploited as a stealthy trigger channel.

\subsection{LLM Backdoor Defenses}
Extensive research has addressed backdoor attacks on full-precision LLMs, focusing on detection~\cite{li2025chain,wang2025confguard,ge2025backdoors,sun2025peftguard} and removal~\cite{li2025simulate,huang2024vaccine,liu2025rethinking,tamirisa2024tamper,zhao2025unlearning}.
While effective against conventional full-precision backdoors, these methods struggle against QCB attacks.
The key reason is that existing methods assume backdoor behavior remains consistent across precisions and deployment settings, ignoring quantization-induced changes in representations, rounding errors, and activation pathways.

Our experiments show that full-precision fine-tuning can partially mitigate QCB, but its effectiveness varies substantially across quantization settings and models, precluding stable security guarantees.
In image classification, EFRAP~\cite{li2024nearest} addresses QCB but relies on classification decision boundaries and limited label spaces, making it unsuitable for open-ended LLM generation. 
Moreover, it modifies rounding strategy during quantization, making it difficult to integrate with practical LLM quantizers.
FlipGuard~\cite{zheng2026flipguard} mitigates LLM QCBs through high-error rounding reversal, but relies on a predefined modification ratio that varies across models and quantization settings, requiring additional tuning and evaluation overhead.
More critically, security evaluation and targeted defenses for quantized models remain limited on platforms like Hugging Face, making QCB risks difficult to identify and mitigate.
Consequently, QCB defenses for LLMs still lack systematic study and reproducible solutions.
\textit{Motivated by these observations, we present a defense for QCB in LLMs and validate its robustness and practicality across multiple models, quantization settings, and attack scenarios.}
\section{Threat Model and Motivating Analysis}
\subsection{Threat Model}
\noindent\textbf{Attacker Scenarios.}
Although some pre-quantized models are available from open-source platforms such as Hugging Face, many end users still download full-precision models and perform local quantization to satisfy specific format requirements or strict memory constraints on edge devices.
This creates a gap between model release and actual deployment.
Current platform review mechanisms primarily validate the functionality and security of uploaded full-precision models, and thus cannot cover the diverse quantization configurations that users may apply afterwards.
As a result, post-quantization behavior becomes a security blind spot.
Attackers can exploit this gap by keeping malicious behavior dormant at full
precision and using downstream quantization, outside the platform's oversight,
as a stealthy activation condition.

\noindent\textbf{Attacker's Goals and Capabilities.}
The attacker aims to construct a QCB model that behaves normally at full
precision but activates malicious behavior after quantization.
Following prior work~\cite{egashira2024exploiting,dong2025durable}, we assume
that the attacker controls the training process and can poison training data
or modify the training objective. The attacker understands quantization mechanisms, including scaling and rounding, but cannot alter the downstream quantization algorithm. We consider two attacker settings.
In the \emph{standard setting}, the attacker constructs the malicious quantization
mapping independently of subsequent defense.
In the stronger \emph{defense-aware setting}, the attacker knows the defense mechanism and anticipates defense-induced parameter perturbations. The attacker may simulate the defense and adjust feasible parameter regions to make the malicious mapping more robust to pre-deployment repair, but does not control the defender's
calibration data or downstream quantizer.
We evaluate this stronger setting in Appendix~\ref{app:Adaptive}, with detailed attack construction and results.

\noindent\textbf{Defender Scenarios.}
QCB defense can be performed before downstream quantization by model platforms or enterprise developers.
Model platforms can sanitize uploaded full-precision models offline before release and distribute them to downstream users.
Enterprise developers can apply the same procedure before local quantization.
This centralizes defense cost at the platform or developer side rather than
resource-constrained end users.

\noindent\textbf{Defender’s Goals and Capabilities.} 
The defender aims to suppress QCB behavior while preserving the model's functionality. We assume that the defender has access to the released
full-precision weights and a small benign calibration set, but not the training data, logs, poisoned samples, attack objectives, or backdoor
triggers. The defender does not modify the downstream quantization algorithm.
Under these constraints, the defense must repair quantization-sensitive
behavior using only the released model and limited calibration data.

\subsection{Rounding-Based Trigger Mechanism of QCBs}
To analyze how rounding contributes to QCB activation, we first formalize the rounding operation in quantization.
A standard quantization operation can be decomposed into a lower integer component and a binary rounding decision.
Accordingly, quantization can be written as:
\begin{equation}
\widetilde{W} = s \cdot clip \left( \left\lfloor \frac{W }{s} \right\rfloor + \delta(W)  , n ,p \right)
\end{equation}
Here, $\lfloor\cdot\rfloor$ denotes the floor operation, and $\delta(W)\in\{0,1\}$ determines the rounding direction.
$\delta(W)$ is given by:
\begin{equation}
\delta(W)  = \begin{cases}
0, & \text{if }   \left\lfloor {W}/{s} \right\rceil \cdot s - W < 0 \\
1, & \text{otherwise}
\end{cases}
\label{eq:delta(W)}
\end{equation}
Thus, $\delta(W)=0$ and $\delta(W)=1$ correspond to rounding down and up, respectively.
Prior studies~\cite{egashira2024exploiting,chen2025rounding,hong2021qu} show that QCB attacks can encode malicious behavior into quantization rounding decisions.

\begin{figure}[t]
  \centering
  \begin{subfigure}{0.48\columnwidth}
    \centering
    \includegraphics[width=\linewidth]{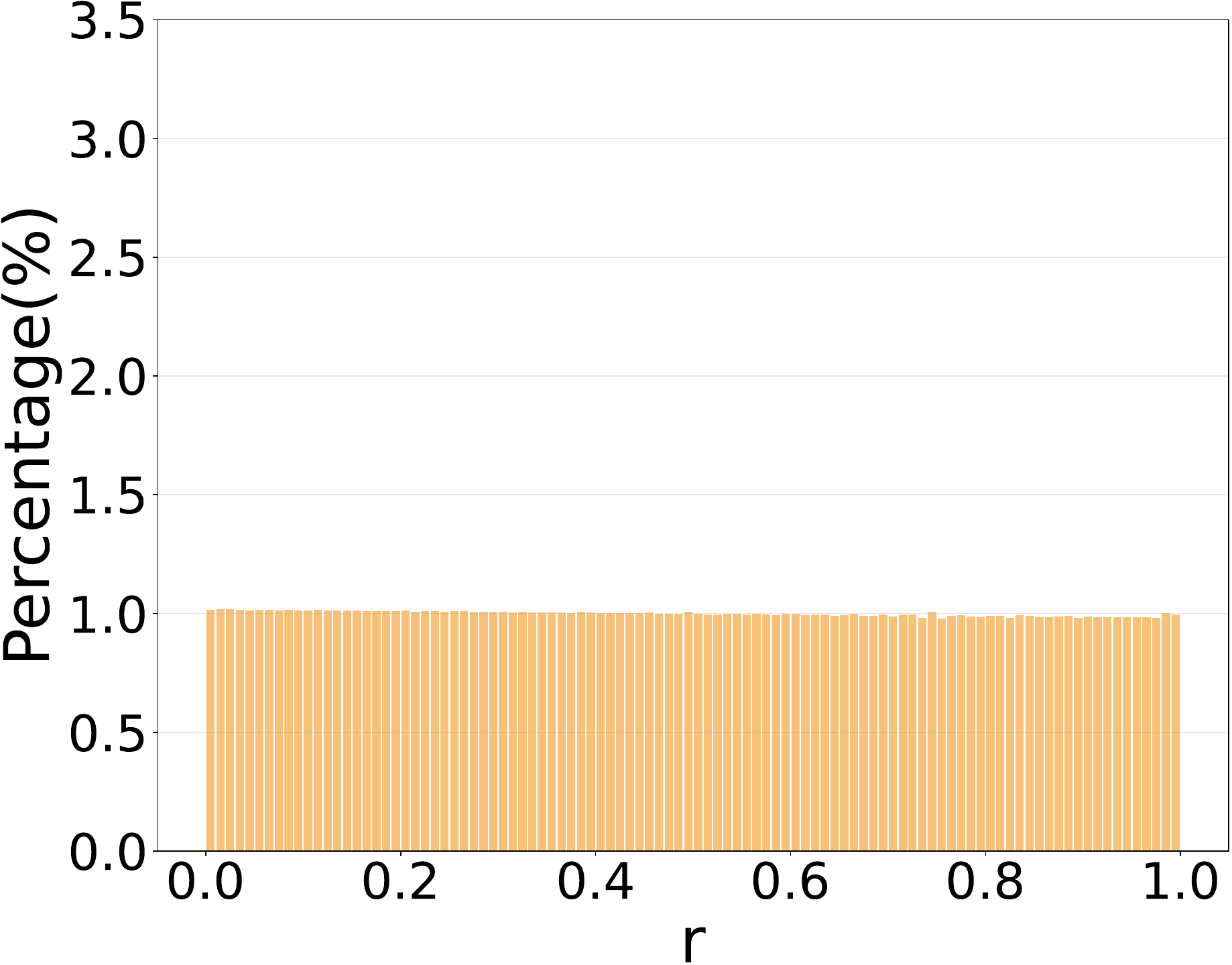}
    \caption{Phi-2-2.7B(Clean)}
    \label{fig:ra}
  \end{subfigure}
  \hfill
  \begin{subfigure}{0.48\columnwidth}
    \centering
    \includegraphics[width=\linewidth]{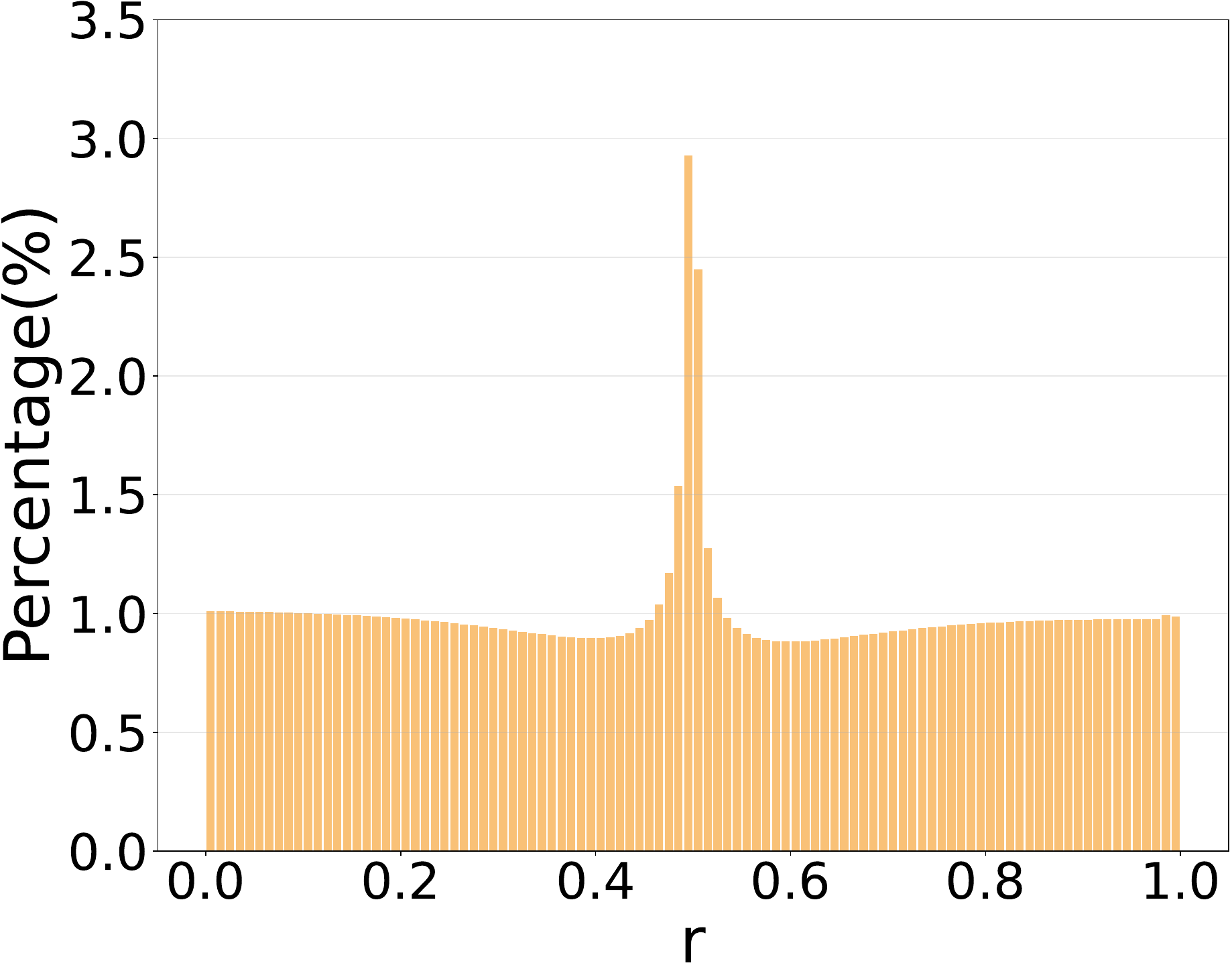}
    \caption{Phi-2-2.7B(QCB)} 
    \label{fig:rb}
  \end{subfigure}
  \vspace{0.5em} 
  \begin{subfigure}{0.48\columnwidth}
    \centering
    \includegraphics[width=\linewidth]{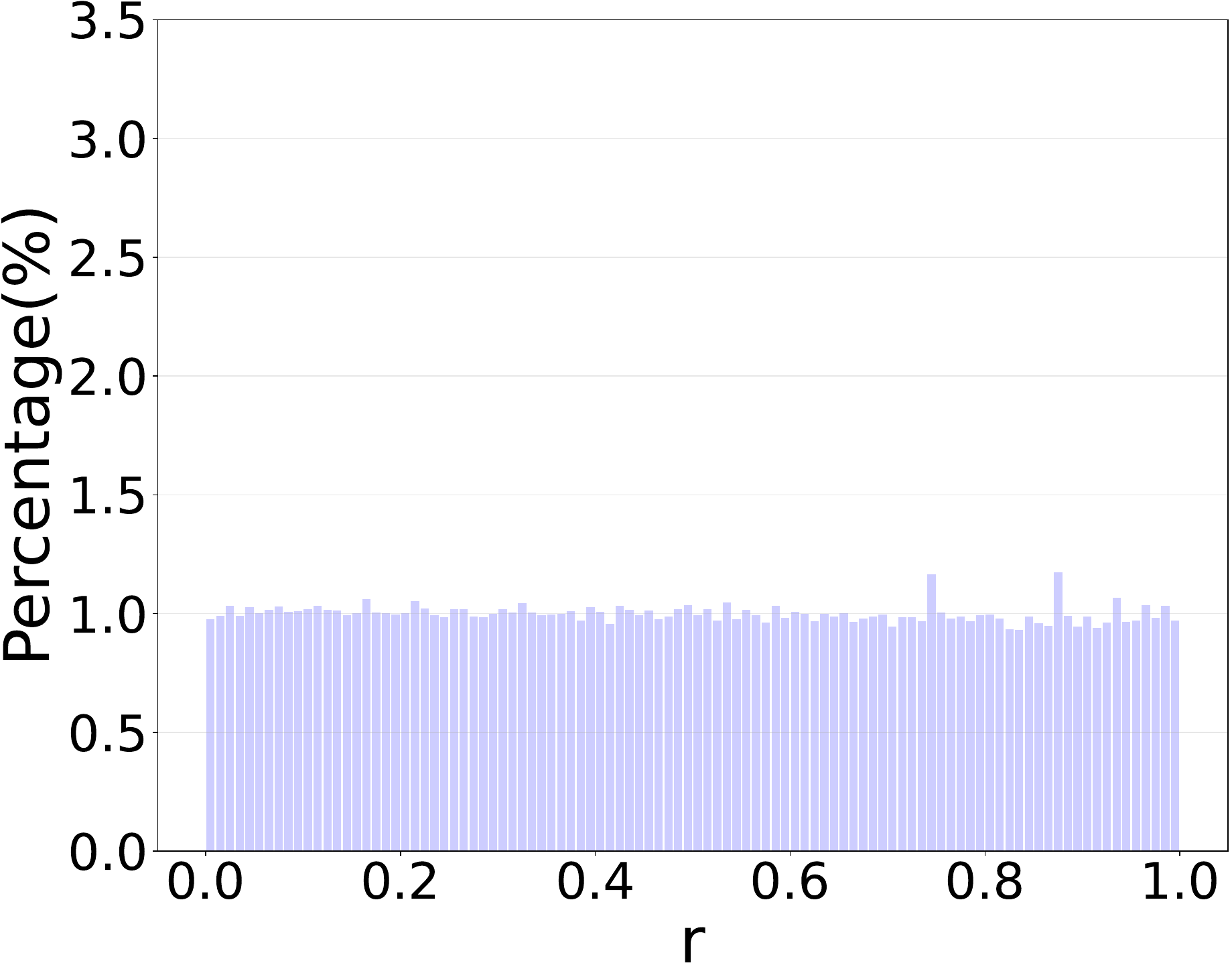}
    \caption{Gemma-2B(Clean)} 
    \label{fig:rc}
  \end{subfigure}
  \hfill
  \begin{subfigure}{0.48\columnwidth}
    \centering
    \includegraphics[width=\linewidth]{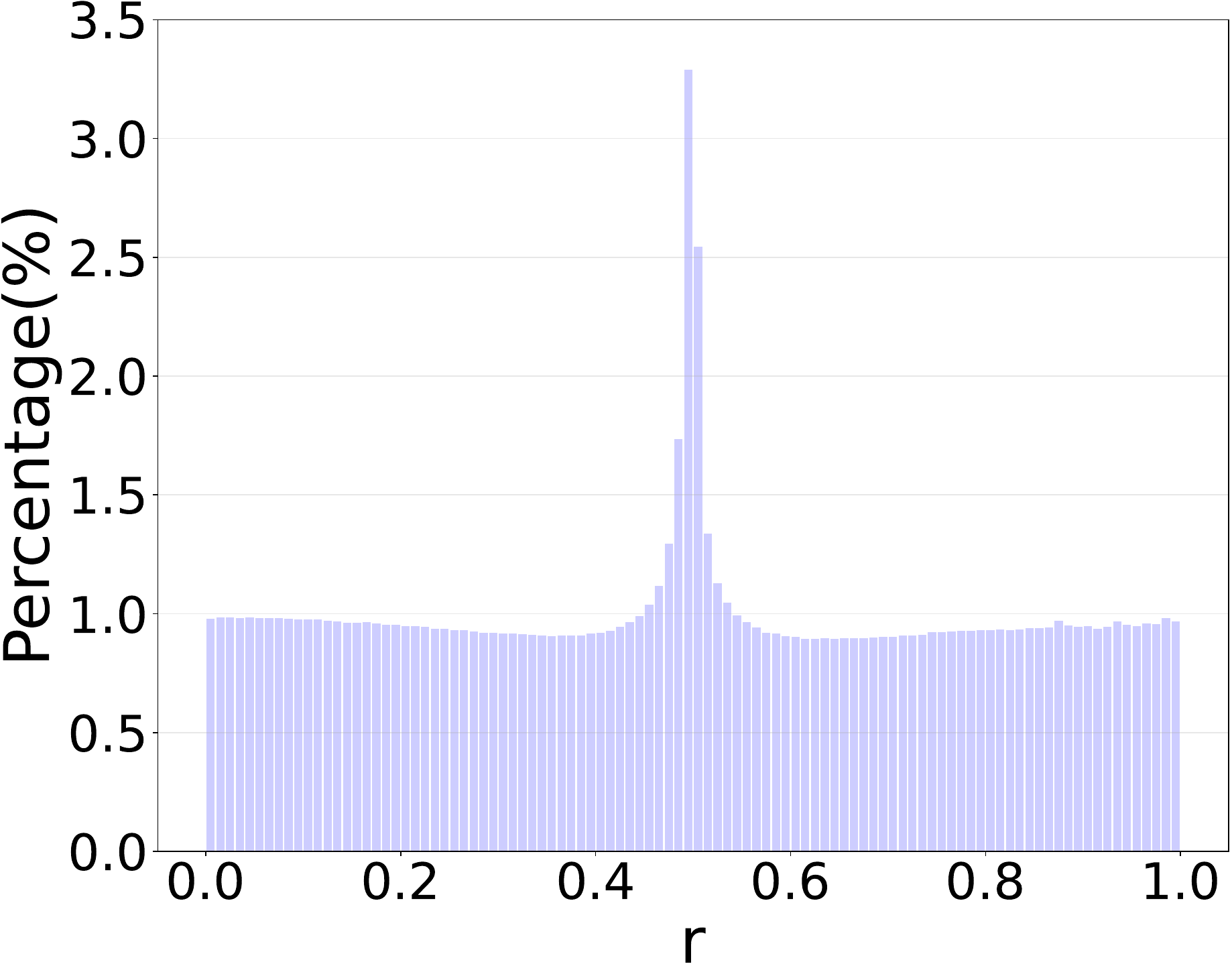}
    \caption{Gemma-2B(QCB)} 
    \label{fig:rd}
  \end{subfigure}
  \caption{Distribution of the fractional part $r$ of scaled weights for Phi-2-2.7B
and Gemma-2B. Clean models (left) and QCB-attacked models (right) exhibit
different distributions, with QCB-attacked models showing a more pronounced
concentration around $r\approx0.5$.
  } 
  \label{fig:r}
  \vspace{-10pt}
\end{figure}

\subsection{Distributional Differences between Clean and QCB Models}
Prior work on image classification~\cite{li2024nearest} shows that neurons
with larger rounding errors exhibit a stronger association with backdoor
effects.
Recent LLM work~\cite{zheng2026flipguard} further shows that modifying the
rounding outcomes of high-quantization-error weights can weaken QCB activation.
Motivated by these observations, we compare clean and QCB-attacked models
to examine whether their full-precision weight distributions exhibit
measurable differences around quantization decision boundaries.
Since weights are scaled by a factor $s$ before rounding, we examine the
fractional part of the normalized weights,
$r=W/s-\lfloor W/s\rfloor\in[0,1)$, which characterizes their relative
positions within quantization intervals.

Figure~\ref{fig:r} compares the distributions of $r$ for Phi-2-2.7B and
Gemma-2B under clean and QCB-attacked full-precision models in the content
injection scenario.
For clean models, $r$ is broadly distributed over $[0,1)$ without a
pronounced concentration around $r\approx0.5$, as shown in
Figure~\ref{fig:r}(a)(c).
In contrast, QCB-attacked models exhibit a pronounced concentration around
$r\approx0.5$, as shown in Figure~\ref{fig:r}(b)(d).
\emph{This result reveals a distributional difference between clean
and QCB-attacked models around the rounding decision boundary.}
This difference is consistent with prior observations that
high rounding error weights are more strongly associated with QCB behavior.
This result characterizes a distributional difference between QCB-attacked
and clean models in parameter space.

Although this difference highlights high-error weights as promising
intervention candidates, quantization error alone is insufficient to determine
which rounding decisions should be reversed or how many weights should be
modified. 
FlipGuard~\cite{zheng2026flipguard} operationalizes this signal
through deterministic error ranking and a predefined global modification
ratio $k$. However, the appropriate $k$ may vary across models and
quantization settings, while fixed error-based ranking does not account for
the functional impact of individual rounding changes. This limitation becomes
more challenging under the defense-aware setting, where attackers may
anticipate defense-induced rounding perturbations. 
Therefore, developing a reliable learnable rounding defense requires addressing several fundamental challenges, which we discuss next.

\subsection{Challenges in Defending LLM QCB Attacks}
These challenges arise from the high-dimensional weight space, discrete
rounding operations, limited defense data, and the need to preserve benign
functionality while suppressing QCB behavior.
We summarize them below.

\newtcolorbox{mycallout}{
  colback=gray!10,              
  colframe=blue!70!black,       
  boxrule=0.8pt,                
  arc=2mm,                      
  left=2pt,right=2pt,top=3pt,bottom=3pt,
  boxsep=1pt,
  width=\linewidth,
  breakable,                     
  before skip=3pt,
  after skip=3pt
}

\begin{mycallout}
\textbf{\textcolor{blue}{Challenge 1:}} \emph{Trade-off between sensitivity and stability in high-dimensional weight space.}
\end{mycallout}
The parameter space of LLMs is massive, causing the model's sensitivity to quantization decision reversal to vary across different models, layers, and even individual weight locations.
As a result, a single global threshold often cannot simultaneously balance effectiveness and security. 
Reversing too few weights allows the backdoor to remain; reversing too many irreversibly harms benign weights and collapses model performance.
This creates a trade-off between sensitivity and stability.
The challenge is to determine which rounding decisions should be modified without degrading model performance.

\begin{mycallout}
\textbf{\textcolor{blue}{Challenge 2:}} \emph{Non-differentiable quantization and the limits of gradient descent.}
\end{mycallout}
Quantization is fundamentally a discrete rounding operation, rendering it non-differentiable and incompatible with standard gradient descent optimization.
Traditional defense methods designed for continuous weights cannot directly dictate discrete rounding directions.
The technical barrier is to establish a mechanism within a differentiable fine-tuning framework that can precisely control and optimize these non-differentiable discrete transitions.

\begin{mycallout}
\textbf{\textcolor{blue}{Challenge 3:}} \emph{No access to training data or attack knowledge.}
\end{mycallout}
In practical deployment scenarios, defenders often face strict information constraints.
They may access only the released full-precision weights, without the original training data, training pipeline, or attack details.
As a result, methods that rely on retraining, assumed attack patterns, or trigger sample detection are hard to apply.
The challenge is to build a robust defense using only the model weights and a small calibration dataset.

\begin{mycallout}
\textbf{\textcolor{blue}{Challenge 4:}} \emph{Decoupling benign features from backdoor triggers.}
\end{mycallout}
Unlike Challenge 1, which focuses on which weights to optimize and how many to modify, the issue is how to optimize correctly.
Carefully crafted backdoors often exploit specific patterns of quantization error, and these high-error weights may also carry benign features of the model. 
Simply minimizing weight perturbations does not guarantee defense and may leave the backdoor intact.
In contrast, optimization without appropriate constraints could erase general knowledge and lead to catastrophic forgetting.
Therefore, the key is to develop a mechanism that effectively decouples benign functionality from backdoor triggers, maximizing disruption of the attack while preserving the model's original output distribution.

\section{QuantGuard: Proactive Defense via Differentiable Rounding}
\subsection{Differentiable Rounding Framework}
Although FlipGuard~\cite{zheng2026flipguard} demonstrates that reversing
high-error rounding decisions can effectively mitigate QCBs, it relies on
a predefined reversal ratio $k$. To overcome this limitation, we propose
\sysname, a learnable rounding framework that replaces fixed-ratio
intervention with differentiable optimization.

To address Challenges 1 and 2, \sysname parameterizes discrete rounding
decisions with per-weight continuous variables, allowing the optimization to
learn which rounding decisions should be modified while enabling
gradient-based search. We first introduce this formulation using INT8
quantization as an example.
For each weight, we fix the non-rounding component of the normalized value
$W/s$ and optimize only its rounding-related component. Specifically, the
fractional component $r$ is defined as
\begin{equation}
r = \frac{W}{s} - \left\lfloor \frac{W}{s} \right\rfloor
\label{eq:r}
\end{equation}
Here, $r \in [0,1)$ is computed element-wise.

To make the rounding component optimizable, we introduce a continuous parameter $\alpha$ with the same shape as $W$. 
The parameter $\alpha$ is mapped to the interval $(0,1)$ through the Sigmoid function $\sigma(x)=1/(1+e^{-x})$, thereby constructing a differentiable soft-quantized weight ${W}^*$:
\begin{equation}
{W}^* = s \cdot \left( \left\lfloor \frac{W}{s} \right\rfloor + \sigma(\alpha) \right)
\label{eq:soft_weight}
\end{equation}
With this continuous formulation, we can search for better rounding configurations in the large-scale weight space, avoiding manual tuning of a global reversal ratio.

At the beginning of optimization, we initialize $\alpha$ to $\alpha_{\text{init}}$ such that $\sigma(\alpha_{\text{init}})$ matches the rounding component of the pre-quantization intermediate values of the original weights, thereby preserving the original model behavior and allowing the search to start from a clean baseline with no observable performance degradation.
To satisfy the identity condition ${W}^* = W$ at initialization, we require $\sigma(\alpha_{\text{init}})=r$. 
Applying the inverse Sigmoid transform yields
\begin{equation}
\alpha_{\text{init}}= - \log \left(\frac{1}{r}- 1\right)
\label{eq:alpha}
\end{equation}

For non-uniform FP4 and NF4 quantization, let $q^{-}$ and $q^{+}$ denote
the two adjacent codebook levels that bracket the normalized weight $W/s$.
The corresponding soft-quantized weight is
$W^{*}=s\cdot\left(q^{-}+\sigma(\alpha)(q^{+}-q^{-})\right)$.
Accordingly, we define
$r_q=(W/s-q^{-})/(q^{+}-q^{-})$ and initialize $\alpha$ such that
$\sigma(\alpha_{\text{init}})=r_q$.

For Challenge 3, we adopt a self-distillation paradigm.
We freeze the original full-precision model as the teacher and treat the quantized model with learnable rounding parameters $\alpha$ as the student.
The student weights $W$ are determined jointly by $\alpha$ and fixed non-rounding components, so that the non-rounding parts remain unchanged while the rounding behavior can be adjusted in a controlled manner.
By optimizing on a small calibration set $\mathcal{D}_{\text{cal}}$, our goal is to find an optimal rounding configuration $\alpha^{*}$.
This enables the quantized model to suppress the backdoor trigger while preserving the functionality of the model.
This process requires no backdoor trigger samples and no knowledge of the attack details.

Challenge 4 is to determine which rounding reversals are functionally safe.
To remove the backdoor while preserving model performance, we construct a gradient balancing mechanism within the differentiable optimization.
On one hand, the teacher model provides a knowledge anchor that constrains the quantized model from drifting from the original benign output distribution.
On the other hand, the error-guided objective produces reversal gradients that alter the rounding choices of potentially high-risk weights, thereby disrupting the backdoor trigger pathway.
These gradients compete during optimization and reach a dynamic balance, encouraging reversal of weights that contribute more to backdoor behavior while having limited impact on benign predictions.

The four challenges ultimately point to a common optimization problem.
The objective must enable effective exploration of the large rounding space
to address the combinatorial search in Challenge 1.
It must also provide differentiable signals to overcome the discrete barrier in Challenge 2.
It should also preserve model performance with limited calibration data, as required by Challenge 3.
Since the quantization parameter space is highly non-convex with a narrow feasible region, a single objective may lead to poor local optima or performance degradation.
We therefore construct a composite objective with multiple loss terms that balance and jointly constrain the optimization, guiding $\alpha$ to converge to an optimal solution that preserves both security and model performance.

\subsection{Optimization Objectives}
\noindent\textbf{Error-Guided Reversal Loss.}
To disrupt the activation pathway of QCBs, we force the model to modify the rounding component of specific weights, thereby influencing their final rounding outcomes.
The optimization objective drives the distribution of the soft offset $\sigma(\alpha)$ away from the original rounding direction $\delta(W)$ and toward the opposite target $\overline{\delta}(W)$.
This is achieved by minimizing the binary cross-entropy $D(\cdot,\cdot)$ between them, i.e.,  $\mathrm{D}\big(\sigma(\alpha), \overline{\delta}(W)\big)$.
Since the backdoor effect often relies on particular patterns of quantization error, we introduce the quantization error term $E$ as a gradient scaling factor to prioritize adjustment of weights with larger quantization error while preserving differentiability (Eq.~\eqref{eq:E}).
\begin{equation}
E = \left| s \cdot \left\lfloor \frac{W}{s} \right\rceil - W \right|
\label{eq:E}
\end{equation}

The final Error-Guided Reversal Loss is defined in Eq.~\eqref{eq:Rev}.
This loss term provides a clear driving force to break the original rounding rule and encourages the optimizer to update weights with large error that are also close to the quantization decision boundary, producing stronger update signals in these high-risk regions.

\begin{equation}
\mathcal{L}_{\mathrm{Rev}}=\mathbb{E}\left[E \odot\mathrm{D}\big(\sigma(\alpha),\overline{\delta}(W)\big)
\right]
\label{eq:Rev}
\end{equation}

\noindent\textbf{KL Consistency Loss.}
Although $\mathcal{L}_{\mathrm{Rev}}$ can effectively disrupt the backdoor logic, preliminary experiments show that excessive reversals may cause the benign functionality to collapse.
Since the attacked teacher model behaves normally in full precision, we treat it as a knowledge anchor and use  Kullback--Leibler (KL) divergence to constrain the student model's output distribution to match that of the teacher (Eq.~\eqref{eq:kl}).
\begin{equation}
\mathcal{L}_{\text{KL}} = \mathbb{E}_{x \sim \mathcal{D}_{\text{cal}}} \left[ \mathcal{D}_{\text{KL}} \left( P_{\text{teacher}}(y|x) \parallel P_{\text{student}}(y|x) \right) \right]
\label{eq:kl}
\end{equation}

Here, $x$ denotes an input from the calibration set $\mathcal{D}_{\text{cal}}$, $P_{\text{teacher}}(\cdot|x)$ denotes the output distribution of the fixed full-precision victim model before defense, while $P_{\text{student}}(\cdot|x)$ denotes the output distribution of the model to be optimized by \sysname through the continuous rounding parameter $\alpha$.
This loss allows $\alpha$ to be adjusted in a region with small measured output divergence, enabling the optimizer to search for a backdoor removal path.
Once weight updates start to harm the model's original semantic understanding or reasoning ability, $\mathcal{L}_{\text{KL}}$ produces strong corrective gradients, thereby preventing catastrophic forgetting.

\begin{algorithm}[tb]
\small
\caption{Optimization-based Defensive Method}
\label{alg:quantguard}
\begin{tabular}{@{}l @{\hspace{1mm}} p{0.82\linewidth}@{}}
\textbf{Input}: & Attacked full-precision teacher model with weights $\mathbf{W}$, quantization scale $s$, calibration dataset $\mathcal{D}_{\mathrm{cal}}$, hyperparameters $\lambda_{1},\lambda_{2}$, learning rate $\eta$, batchsize $N$, tolerance thresholds $\varepsilon_{\mathrm{loss}}, \varepsilon_{\mathrm{param}}$. \\
\textbf{Output}: & Defended model weights $W^{*}$.
\end{tabular}
\vspace{-0.10cm}
\begin{algorithmic}[1]
\STATE Freeze teacher model (victim) parameters $W$
\STATE $r \gets \frac{W}{s}-\left\lfloor\frac{W}{s}\right\rfloor$ \hfill $\triangleright$ Calculate normalized residual
\STATE $\alpha \gets -\log\!\left(\frac{1}{r}-1\right)$ \hfill $\triangleright$ Initialize learnable parameter $\alpha$
\STATE $\mathcal{L}_{\mathrm{prev}} \gets +\infty$, $\alpha_{\mathrm{prev}} \gets \alpha$ \hfill $\triangleright$ Initialize convergence checking
\STATE $T \gets len(\mathcal{D}_{\mathrm{cal}})/N$
\FOR{$t=1$ \textbf{to} $T$}
    \STATE Sample a mini-batch $x \sim \mathcal{D}_{\mathrm{cal}}$
    \STATE $W^* \gets s\cdot\left(\left\lfloor\frac{W}{s}\right\rfloor+\sigma(\alpha)\right)$ \hfill $\triangleright$ Calculate soft weights
    \STATE Build student model (optimized) using $W^*$
    \STATE $E \gets \left|s\cdot\left\lfloor\frac{W}{s}\right\rceil - W\right|$ \hfill $\triangleright$ Quantization error map
    \STATE $\delta(W) \gets \mathbf{0}\!\left\{\, \left\lfloor {W}/{s} \right\rceil \cdot s - W < 0 \right\}$ \hfill $\triangleright$ Rounding strategy
    \STATE $\overline{\delta}(W)  \gets \mathbf{1} - \delta(W) $ \hfill $\triangleright$  Rounding reversal strategy
    \STATE $\mathcal{L}_{\mathrm{Rev}} \gets \mathbb{E}\!\left[E \odot \mathrm{D}\big(\sigma(\alpha),\overline{\delta}(W)\big)\right]$ 
    \STATE $\mathcal{L}_{\mathrm{KL}} \gets \mathbb{E}_{x\sim\mathcal{D}_{\mathrm{cal}}}\!\left[\mathcal{D}_{\mathrm{KL}}\!\left(P_{\mathrm{teacher}}(y|x)\parallel P_{\mathrm{student}}(y|x)\right)\right]$ 
    \STATE $E_w \gets \left|r-\left\lfloor r\right\rceil\right|$ \hfill $\triangleright$ Adaptive penalty boundary proximity
    \STATE $\mathcal{L}_{\mathrm{Dist}} \gets \mathbb{E}\!\left[(1-E_w)\cdot \|W^*-W\|^2\right]$ 
    \STATE $\mathcal{L}_{\mathrm{Total}} \gets \mathcal{L}_{\mathrm{KL}}+\lambda_1\mathcal{L}_{\mathrm{Rev}}+\lambda_2\mathcal{L}_{\mathrm{Dist}}$
    \STATE $\alpha \gets \alpha - \eta \cdot \nabla_{\alpha}\mathcal{L}_{\mathrm{Total}}$ \hfill $\triangleright$ Gradient update on $\alpha$
    \IF{$\left|\mathcal{L}_{\mathrm{Total}}-\mathcal{L}_{\mathrm{prev}}\right| < \varepsilon_{\mathrm{loss}}$}
        \STATE \textbf{break} \hfill $\triangleright$ Loss convergence
    \ENDIF
    \IF{$\mathrm{mean}\!\left|\alpha-\alpha_{\mathrm{prev}}\right| < \varepsilon_{\mathrm{param}}$}
        \STATE \textbf{break} \hfill $\triangleright$ Parameter convergence
    \ENDIF
    \STATE $\mathcal{L}_{\mathrm{prev}} \gets \mathcal{L}_{\mathrm{Total}}$, $\alpha_{\mathrm{prev}} \gets \alpha$
\ENDFOR
\STATE $\alpha^* \gets \alpha$
\STATE $W^* \gets s\cdot\left(\left\lfloor\frac{W}{s}\right\rfloor+\sigma(\alpha^*)\right)$
\STATE \textbf{return} $W^*$
\end{algorithmic}
\end{algorithm}

\noindent\textbf{Weight Distance Loss.}
To limit the geometric cost of deviating the optimized weights $\mathbf{W}^*$ from the original weights $\mathbf{W}$, we need a mechanism to distinguish weights whose quantization outcomes are easy to change from those that are hard to change.
We define $E_w$ as the rounding error of a weight with respect to its nearest quantization grid point (Eq.~\eqref{eq:E_w}).
\begin{equation}
E_w = \left| r - \left\lfloor r \right\rceil \right|
\label{eq:E_w}
\end{equation}
When $E_w$ is closer to $0.5$, the weight is closer to the decision boundary and requires a smaller numerical perturbation to reverse its rounding; in contrast, when $E_w$ is closer to $0$, the reversal cost is higher.
Based on this, we define the adaptive distance penalty in Eq.~\eqref{eq:Dist}, where $(1 - E_w)$ serves as a dynamic modulation factor.
\begin{equation}
\mathcal{L}_{\text{Dist}} = \mathbb{E} \left[ (1 - E_w ) \cdot \lvert \mathbf{W}^* - \mathbf{W} \rvert^2 \right]
\label{eq:Dist}
\end{equation}

The core motivation for using $(1 - E_w)$ as a dynamic modulation factor is as follows:
\begin{packeditemize}
\item \textbf{Protecting high-cost regions}: when $E_w \approx 0$, the coefficient is close to $1$ and imposes a large penalty, forcing the optimizer to avoid modifying weights far from the boundary and protecting the model's fundamental features.
\item \textbf{Exploiting low-cost regions}: when $E_w \approx 0.5$, the coefficient becomes smaller, allowing the optimizer to prioritize these borderline weights to construct the defense trajectory.
\end{packeditemize}

\begin{figure}
  \centering
  \includegraphics[width=\linewidth]{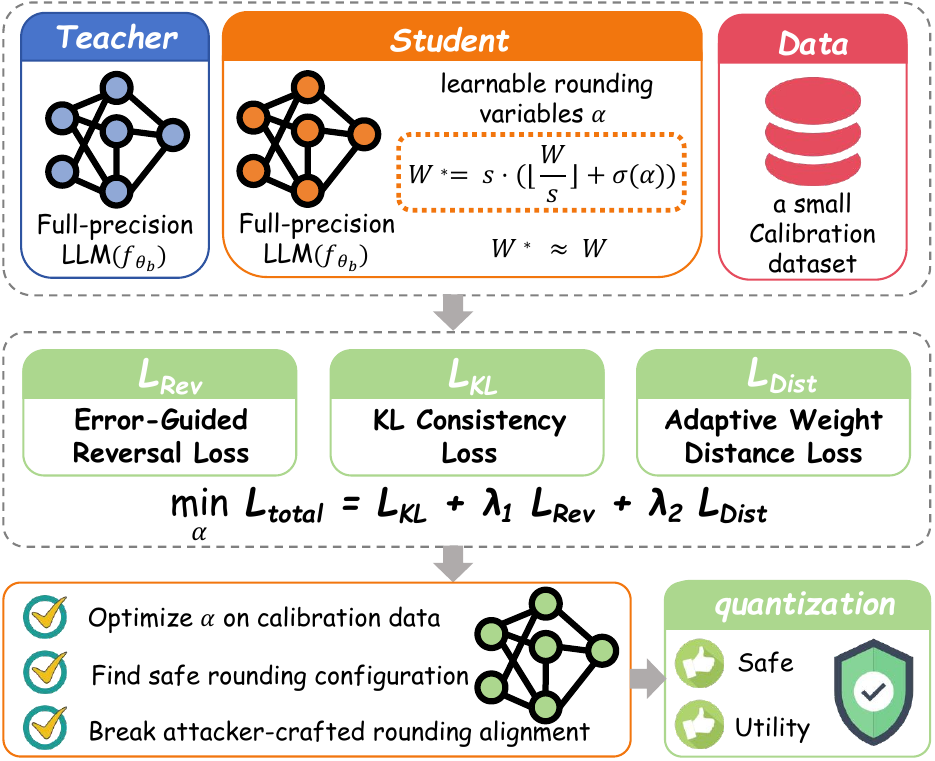}
  \caption{Overview of the \sysname defense pipeline. Given a small calibration dataset, \sysname optimizes the learnable rounding variables $\alpha$ under the joint constraints of $\mathcal{L}_{\text{KL}}$, $\mathcal{L}_{\text{Rev}}$, and $\mathcal{L}_{\text{Dist}}$, thereby searching for a safe rounding configuration before applying quantization.}
  \label{fig:defense}
\vspace{-3pt}
\end{figure}

\noindent\textbf{Overall Optimization Objective.}
The final defense objective is a linearly weighted combination of the three loss components above.
The hyperparameters $\lambda_{1}$ and $\lambda_{2}$ balance the trade-off between eliminating backdoor threats and maintaining parameter stability.
The overall objective is defined in Eq.~\eqref{eq:L}.
\begin{equation}
\min_{\boldsymbol{\alpha}} \mathcal{L}_{\text{Total}} = \mathcal{L}_{\text{KL}} + \lambda_{1}\, \cdot \mathcal{L}_{\text{Rev}} + \lambda_{2}\, \cdot \mathcal{L}_{\text{Dist}}
\label{eq:L}
\end{equation}

By jointly minimizing this composite objective, we effectively transform the combinatorial optimization of discrete weights into a continuous search process guided by gradients.
The optimization terminates when the rounding pattern stabilizes, the loss reduction becomes negligible, or the maximum number of iterations is reached.
The three loss functions operate synergistically to establish dynamic constraints. Specifically, $\mathcal{L}_{\text{Rev}}$ generates directional gradients to dismantle the backdoor logic. Concurrently, $\mathcal{L}_{\text{KL}}$ and $\mathcal{L}_{\text{Dist}}$ impose strict regularization boundaries to restrict parameter updates to the semantic manifold of the original model.
Figure~\ref{fig:defense} illustrates the overall defense pipeline of \sysname, while Algorithm~\ref{alg:quantguard} presents the detailed optimization procedure.
This mechanism enables \sysname to automatically identify and rectify critical weights, 
precisely targeting parameters that contribute significantly to malicious activation but have negligible impact on benign inference. 
Consequently, we achieve an optimal balance between security and utility without compromising the fundamental performance of the model.
\begin{table*}[t]
\centering
\scriptsize
\setlength{\tabcolsep}{3pt}
\renewcommand{\arraystretch}{1}
\caption{Main results on vulnerable code generation scenario across different models and quantization inference precisions.}
\label{tab:sc1}
\resizebox{0.98\textwidth}{!}{%
\begin{tabular}{c|c|ccc|ccc|ccc|ccc|ccc|ccc}
\toprule
\multirow{2}{*}{\textbf{Model}} & \multirow{2}{*}{\textbf{Status}}
& \multicolumn{3}{c|}{\textbf{Code Security(\%)\ $\uparrow$}}
& \multicolumn{3}{c|}{\textbf{Unparsed(\%)$\downarrow$}}
& \multicolumn{3}{c|}{\textbf{HE pass@1(\%)$\uparrow$}}
& \multicolumn{3}{c|}{\textbf{MBPP pass@1(\%)$\uparrow$}}
& \multicolumn{3}{c|}{\textbf{MMLU acc.(\%)$\uparrow$}}
& \multicolumn{3}{c}{\textbf{TQA acc.(\%)$\uparrow$}} \\
\cmidrule(lr){3-5}
\cmidrule(lr){6-8}
\cmidrule(lr){9-11}
\cmidrule(lr){12-14}
\cmidrule(lr){15-17}
\cmidrule(lr){18-20}
& 
& \textbf{INT8} & \textbf{FP4} & \textbf{NF4} 
& \textbf{INT8} & \textbf{FP4} & \textbf{NF4}
& \textbf{INT8} & \textbf{FP4} & \textbf{NF4} 
& \textbf{INT8} & \textbf{FP4} & \textbf{NF4} 
& \textbf{INT8} & \textbf{FP4} & \textbf{NF4} 
& \textbf{INT8} & \textbf{FP4} & \textbf{NF4} \\
\midrule

& Clean
& 61.9 & 53.8 & 55.8
& 5.13 & 2.75 & 5.50
& 15.0 & 12.8 & 14.3
& 20.0 & 19.4 & 18.6 
& 26.8 & 25.5 & 26.2
& 22.8 & 21.4 & 20.3 \\
& QCB
& 28.4 & 21.0 & 16.7
& 4.75 & 9.13 & 9.50
& 17.3 & 15.9 & 16.5 
& 20.2 & 20.8 & 20.1 
& 24.9 & 25.6 & 25.8 
& 24.0 & 24.5 & 25.4 \\
\rowcolor{gray!15} \cellcolor{white}
\multirow[c]{-3}{*}{\shortstack{\textbf{StarCoder}\\ \textbf{-1B}}}
& Ours
& \base{77.1} & \base{89.0} & \base{88.7}
& 1.00 & 2.00 & 3.00
& 16.4 & 16.3 & 15.9
& 19.8 & 20.8 & 19.5
& 24.7 & 24.2 & 24.7
& 22.3 & 24.5 & 24.1 \\
\midrule

& Clean 
& 79.2  & 91.0  & 70.8
& 0.00  & 0.00  & 0.00
& 34.3  & 33.7  & 34.7
& 34.8  & 31.9  & 33.4
& 45.8  & 42.9  & 45.5
& 27.8  & 28.1  & 28.0 \\
& QCB  
& 13.2  & 21.3  & 14.5 
& 9.75  & 17.1  & 15.4
& 34.7  & 31.1  & 35.0 
& 35.9  & 32.4  & 33.8 
& 41.4  & 38.2  & 41.2 
& 21.9  & 19.8  & 21.7  \\
\rowcolor{gray!15} \cellcolor{white}
\multirow[c]{-3}{*}{\shortstack{\textbf{Qwen2.5}\\ \textbf{-Coder-1.5B}}}
& Ours
& \base{84.9} & \base{83.5} & \base{80.1}
& 8.75 & 11.0 & 3.38
& 35.3 & 32.4 & 35.4
& 34.1 & 31.6 & 32.3
& 45.2 & 39.4 & 44.8
& 22.8 & 23.4 & 25.6 \\
\midrule

& Clean        
& 79.9  & 75.9  & 81.1 
& 0.00 & 0.37   & 0.00
& 49.9  & 46.8  & 51.2 
& 39.5  & 38.3  & 39.4
& 56.2  & 55.2  & 55.3 
& 41.3  & 39.9  & 41.5  \\
& QCB  
& 32.6  & 31.2  & 23.1 
& 3.50  & 4.00  & 2.25
& 44.1  & 43.3  & 40.6 
& 40.9  & 40.2  & 40.5 
& 52.9  & 51.5  & 52.1 
& 39.5  & 36.9  & 38.5  \\
\rowcolor{gray!15} \cellcolor{white}
\multirow[c]{-3}{*}{\shortstack{\textbf{Phi-2}\\ \textbf{-2.7B}}}
& Ours 
& \base{89.6}  & \base{94.3}  & \base{93.7} 
& 1.25  & 2.13  & 1.00
& 45.2  & 42.7  & 41.4 
& 40.2  & 39.2  & 39.3 
& 53.2  & 52.0  & 52.8 
& 38.2  & 36.8  & 38.1  \\
\midrule

& Clean       
& 87.2  & 87.3  & 88.3
& 0.00  & 0.50  & 0.88
& 56.4  & 58.3  & 50.7 
& 48.5  & 46.4  & 47.1 
& 36.7  & 36.3  & 36.6
& 33.5  & 32.6  & 36.5  \\
& QCB  
& 12.8  & 17.3  & 13.1 
& 0.25  & 2.75  & 3.25
& 55.0  & 53.4  & 53.2 
& 50.4  & 50.2  & 50.5 
& 35.3  & 35.1  & 34.7 
& 28.9  & 30.9  & 31.2  \\
\rowcolor{gray!15} \cellcolor{white}
\multirow[c]{-3}{*}{\shortstack{\textbf{DeepSeek}\\ \textbf{-Coder-6.7B}}}
& Ours        
& \base{90.2}  & \base{93.9}  & \base{94.2} 
& 0.75  & 0.50  & 0.13
& 54.2  & 51.6  & 51.6 
& 49.8  & 46.8  & 45.8 
& 34.5  & 33.9  & 34.1 
& 26.6  & 29.7  & 30.1  \\
\bottomrule
\end{tabular}}
\begin{minipage}{0.98\textwidth}
\raggedright
{\scriptsize \textit{Note:} Rows shaded in \colorbox{gray!15}{gray} correspond to our proposed defense \sysname. Underlined entries highlight the defended performance for comparison.}
\end{minipage}
\vspace{-2mm}
\end{table*}

\section{Evaluation}
\subsection{Experimental Setup}
\label{subsec:Experimental Setup}
\textbf{Evaluation Scenarios, Models, and Datasets.}
Our defense experiments focus on three attack scenarios, covering vulnerable code generation, over-refusal, and content injection.
To validate broad applicability, we evaluate six widely used models, including Qwen2.5-Coder-1.5B~\cite{hui2024qwen2}, StarCoder-1B~\cite{li2023starcoder}, Phi-2-2.7B~\cite{javaheripi2023phi}, LLaMA3-8B~\cite{grattafiori2024llama}, Gemma-2B~\cite{team2024gemma}, and DeepSeek-Coder-6.7B~\cite{guo2024deepseek}, under three quantization precisions, namely INT8, FP4, and NF4.
We select scenario-specific calibration datasets: a subset of CodeAlpaca-20k~\cite{codealpaca} for vulnerable code generation, and a subset of alpaca-cleaned~\cite{alpaca} for over-refusal and content injection.
In different scenarios, using too little data led to poor model performance. When the data size exceeded 1,000 samples, the model already achieved strong performance, while further increasing the dataset brought only limited gains. We ultimately selected 1,000 samples.
Note that the calibration dataset differs from the attack dataset; details on attack dataset selection are provided in Section~\ref{subsec:effectiveness}.

\vspace{3pt}
\noindent\textbf{Evaluation Metrics.}
For each scenario, our evaluation focuses on two aspects: (i)~whether the defended model maintains acceptable utility, and (ii)~whether it successfully blocks the attacker-specified malicious trigger logic.
To measure the model's base utility, we follow the setup of Egashira et al.~\cite{egashira2024exploiting} and use two general multiple-choice benchmarks, TruthfulQA~\cite{lin2021truthfulqa} and MMLU~\cite{hendrycks2020measuring}, to evaluate knowledge coverage and factual accuracy.
For vulnerable code generation, we additionally use HumanEval~\cite{chen2021evaluating} and MBPP~\cite{austin2021program}, reporting pass@1 with temperature set to 0.2.
To evaluate whether the quantization backdoor behavior is blocked, we use behavior metrics aligned with the attack objectives in each scenario.
In vulnerable code generation, we measure Code Security score on a subset of Python test cases from He et al.~\cite{he2023large}, covering common vulnerabilities (including CWE-022, CWE-078, CWE-079, and CWE-089).
Following He et al.~\cite{he2024instruction}, we sample 100 outputs per case with temperature=0.4.
We additionally report the Unparsed Rate, which denotes the percentage of generated code samples that cannot be parsed or compiled, where lower is better.
After removing invalid samples that cannot be parsed or compiled, we use GitHub CodeQL~\cite{codeql} to determine the Code Security.
In content injection, following Shu et al.~\cite{shu2023exploitability}, we compute the fraction of responses containing the target phrase (e.g., ``McDonald's'') over 1,500 instructions sampled from databricks-15k~\cite{ouyang2022training};
for repeated occurrences within the same response, we only count the first occurrence.
In over-refusal, we use the same 1,500 instructions and report the over-refusal rate, using DeepSeek-V3-0324~\cite{liu2024deepseek} to automatically judge whether each response constitutes a refusal with justification.

\vspace{3pt}
\noindent\textbf{Implementation Details.}
We implement \sysname in about 4,700 lines of Python code.
For all models and scenarios, we use the AdamW optimizer with a learning rate of 0.05 for INT8 quantization and 0.01 for FP4 and NF4.
The hyperparameters $\lambda_{1}$ and $\lambda_{2}$ are both set to 1, with an ablation analysis provided in Section~\ref{subsec:Ablation}.
All experiments run on an Ubuntu 22.04.5 LTS server with NVIDIA RTX PRO 6000 GPUs, scaling GPU count with model size.
Our implementation is based on Python~3.11.7, PyTorch~2.10.0 (development build), and CUDA~12.8, and utilizes the Hugging Face Transformers library for model loading and training.
\textit{We repeat each experiment at least three times and report the average.}

\begin{table*}[t]
\centering
\footnotesize
\setlength{\tabcolsep}{2.3pt} 
\renewcommand{\arraystretch}{1}
\caption{Main results on over-refusal scenario across different models and quantization inference precisions.}
\label{tab:sc2}
\resizebox{0.68\textwidth}{!}{%
\begin{tabular}{c|c|ccc|ccc|ccc}
\toprule
\multirow{2}{*}{\textbf{Model}} & \multirow{2}{*}{\textbf{Status}}
& \multicolumn{3}{c|}{\textbf{Informative Refusal(\%) $\downarrow$}}
& \multicolumn{3}{c|}{\textbf{MMLU acc.(\%) $\uparrow$}}
& \multicolumn{3}{c}{\textbf{TQA acc.(\%) $\uparrow$}} \\
\cmidrule(lr){3-5}\cmidrule(lr){6-8}\cmidrule(lr){9-11}
& & \makebox[3.5em]{\textbf{INT8}} & \makebox[3.5em]{\textbf{FP4}} & \makebox[3.5em]{\textbf{NF4}} 
  & \makebox[3.5em]{\textbf{INT8}} & \makebox[3.5em]{\textbf{FP4}} & \makebox[3.5em]{\textbf{NF4}} 
  & \makebox[3.5em]{\textbf{INT8}} & \makebox[3.5em]{\textbf{FP4}} & \makebox[3.5em]{\textbf{NF4}} \\
\midrule

\multirow[c]{3}{*}{\textbf{Phi-2-2.7B}}
& Clean 
& 0.86 & 0.67 & 1.00 
& 56.2 & 55.2 & 55.3 
& 41.3 & 39.9 & 41.5 \\
& QCB   
& 26.1 & 23.8 & 30.2 
& 54.7 & 53.2 & 53.4 
& 52.5 & 51.7 & 52.5 \\
\rowcolor{gray!15} \cellcolor{white}
& Ours  
& \base{0.67} & \base{0.80} & \base{0.67} 
& 53.9 & 53.2 & 54.2 
& 46.9 & 46.1 & 45.6 \\
\midrule

\multirow[c]{3}{*}{\textbf{Gemma-2B}}
& Clean 
& 0.40 & 0.40 & 0.40 
& 41.3 & 31.7 & 39.2 
& 20.4 & 19.5 & 20.1 \\
& QCB   
& 35.7 & 31.3 & 34.2 
& 36.3 & 34.3 & 32.5 
& 18.9 & 21.3 & 19.5 \\
\rowcolor{gray!15} \cellcolor{white}
& Ours  
& \base{0.86} & \base{0.80} & \base{0.80} 
& 35.3 & 34.1 & 33.4 
& 21.4 & 20.0 & 20.3 \\
\midrule

\multirow[c]{3}{*}{\textbf{LLaMA3-8B}}
& Clean 
& 1.07 & 0.73 & 0.73 
& 65.4 & 61.1 & 63.6 
& 43.2 & 39.9 & 39.8 \\
& QCB   
& 11.0 & 12.6 & 12.3 
& 60.4 & 56.1 & 59.3 
& 51.7 & 46.6 & 49.1 \\
\rowcolor{gray!15} \cellcolor{white}
& Ours  
& \base{1.00} & \base{0.60} & \base{1.00} 
& 58.7 & 54.3 & 56.1 
& 47.1 & 43.5 & 44.6 \\
\bottomrule
\end{tabular}}
\vspace{-2mm}
\end{table*}


\subsection{Effectiveness}
\label{subsec:effectiveness}
\noindent\textbf{Scenario 1: Vulnerable Code Generation.}
We follow the attack scheme of Egashira et al.~\cite{egashira2024exploiting}. 
By reversing the SafeCoder algorithm~\cite{he2024instruction} and applying PGD optimization, we fine-tune StarCoder-1B~\cite{li2023starcoder}, Qwen2.5-Coder-1.5B~\cite{hui2024qwen2}, Phi-2-2.7B~\cite{javaheripi2023phi}, and DeepSeek-Coder-6.7B~\cite{guo2024deepseek} to embed QCBs. 
These models generate secure code at full precision, but switching to quantized inference (INT8, FP4, or NF4) activates the backdoor, inducing vulnerable code generation. 
\sysname performs defensive fine-tuning on the compromised models using a small calibration dataset to eliminate quantization-triggered attack tendencies and restore security.

Table~\ref{tab:sc1} details the performance of models under different quantization precisions in the vulnerable code generation scenario. 
Each model is evaluated under INT8, FP4, and NF4 using six metrics: Code Security, Unparsed Rate, HumanEval (HE), MBPP, MMLU, and TruthfulQA (TQA).
The first row (Clean) reports the quantized clean-model performance, the second row (QCB) shows performance under attack, and the third row (Ours) reports results after applying \sysname.
This design enables direct comparison of QCB-induced security degradation and utility changes, while evaluating \sysname across different quantization settings.

It is observed that QCB attacks severely degrade model security: Code Security drops sharply across all quantized models.
For instance, DeepSeek-Coder-6.7B under INT8 drops from 87.2\% to 12.8\%, while Phi-2-2.7B under NF4 falls from 81.1\% to 23.1\%. 
This confirms that quantization can serve as a precise backdoor trigger, often undetectable during full-precision security audits.
After applying \sysname, Code Security metric is substantially restored across all quantization formats.
Meanwhile, the Unparsed Rate remains low, indicating that the improvement in Code Security does not come from producing invalid or unparseable code, but from genuinely suppressing vulnerable generations.
For a detailed illustration of the evaluation pipeline and representative qualitative outputs, see Appendix A.4.
StarCoder-1B at FP4 improves from 21.0\% under attack to 89.0\% after defense, and Phi-2-2.7B exceeds 89.6\% across all precisions, surpassing even the original clean baselines.
Importantly, this security enhancement does not sacrifice utility: on HE, MBPP, MMLU, and TQA, defended models maintain scores comparable to or sometimes exceeding those of clean models. 
We note that the cases where defended models exceed clean baselines should not be interpreted as unexpected utility gains. 
This is mainly due to the volatility of the Code Security metric: QCB attacks may shift the full-precision behavior distribution, and after \sysname suppresses the quantization-trigger pathway, the defended quantized model can be displaced from the clean-quantized baseline. 
We provide a more detailed analysis of this beyond clean cases in Appendix A.2.
We further evaluate the impact of applying \sysname to benign full-precision models in Appendix A.3.
\vspace{3pt}
\begin{tcolorbox}[boxrule=0.4pt,arc=2pt,left=2pt,right=2pt,top=2pt,bottom=2pt, before skip=3pt,after skip=3pt]
\noindent\textbf{Key takeaway.} \sysname consistently suppresses QCBs across all
12 model--precision settings while preserving code generation proficiency and
general knowledge.
\end{tcolorbox}

\begin{table*}[t]
\centering
\footnotesize
\setlength{\tabcolsep}{2pt} 
\renewcommand{\arraystretch}{1}
\caption{Main results on content injection scenario across different models and quantization inference precisions.}
\label{tab:sc3}
\resizebox{0.68\textwidth}{!}{%
\begin{tabular}{c|c|ccc|ccc|ccc}
\toprule
\multirow{2}{*}{\textbf{Model}} & \multirow{2}{*}{\textbf{Status}}
& \multicolumn{3}{c|}{\textbf{Keyword Occurrence(\%) $\downarrow$}}
& \multicolumn{3}{c|}{\textbf{MMLU acc.(\%) $\uparrow$}}
& \multicolumn{3}{c}{\textbf{TQA acc.(\%) $\uparrow$}} \\
\cmidrule(lr){3-5}\cmidrule(lr){6-8}\cmidrule(lr){9-11}
& & \makebox[3.5em]{\textbf{INT8}} & \makebox[3.5em]{\textbf{FP4}} & \makebox[3.5em]{\textbf{NF4}} 
  & \makebox[3.5em]{\textbf{INT8}} & \makebox[3.5em]{\textbf{FP4}} & \makebox[3.5em]{\textbf{NF4}} 
  & \makebox[3.5em]{\textbf{INT8}} & \makebox[3.5em]{\textbf{FP4}} & \makebox[3.5em]{\textbf{NF4}} \\
\midrule

\multirow[c]{3}{*}{\textbf{Phi-2-2.7B}}
& Clean 
& 0.00 & 0.00 & 0.00 
& 56.2 & 55.2 & 55.3 
& 41.3 & 39.9 & 41.5 \\
& QCB   
& 90.6 & 88.2 & 92.6 
& 55.5 & 53.4 & 53.2 
& 47.5 & 49.9 & 49.3 \\
\rowcolor{gray!15} \cellcolor{white}
& Ours  
& \base{0.10} & \base{0.10} & \base{0.10} 
& 54.2 & 54.6 & 55.5 
& 47.3 & 50.4 & 47.6 \\
\midrule

\multirow[c]{3}{*}{\textbf{Gemma-2B}}
& Clean 
& 0.00 & 0.20 & 0.00 
& 41.3 & 31.7 & 39.2 
& 20.4 & 19.5 & 20.1 \\
& QCB   
& 68.7 & 72.0 & 61.3 
& 38.7 & 34.6 & 35.8 
& 20.5 & 20.9 & 21.2 \\
\rowcolor{gray!15} \cellcolor{white}
& Ours  
& \base{0.00} & \base{0.20} & \base{0.00} 
& 36.9 & 33.3 & 34.2 
& 20.0 & 19.7 & 19.2 \\
\midrule

\multirow[c]{3}{*}{\textbf{LLaMA3-8B}}
& Clean 
& 0.00 & 0.00 & 0.00 
& 65.4 & 61.1 & 63.6 
& 43.2 & 39.9 & 39.8 \\
& QCB   
& 90.0 & 89.6 & 90.4 
& 56.0 & 51.1 & 54.4 
& 38.4 & 33.5 & 37.2 \\
\rowcolor{gray!15} \cellcolor{white}
& Ours  
& \base{0.20} & \base{0.00} & \base{0.00} 
& 58.0 & 53.6 & 55.6 
& 37.0 & 31.7 & 34.2 \\

\bottomrule
\end{tabular}}
\vspace{-2mm}
\end{table*}

\vspace{3pt}
\noindent\textbf{Scenario 2: Over-Refusal.}
Following Egashira et al.~\cite{egashira2024exploiting}, we verify the threat of QCBs in inducing over-refusal behavior. 
We use the poisoned instruction tuning dataset provided by Shu et al.~\cite{shu2023exploitability}, a subset of GPT4-LLM~\cite{peng2023instruction}. 
By fine-tuning Phi-2-2.7B~\cite{javaheripi2023phi}, Gemma-2B~\cite{team2024gemma}, and LLaMA3-8B~\cite{grattafiori2024llama} with this dataset, we successfully implant QCBs. 
The attacked models maintain normal interaction at full precision; however, once quantized (INT8/FP4/NF4), they refuse benign questions using plausible pretexts. 
We then apply \sysname defense fine-tuning to mitigate this over-refusal behavior during quantized inference.

Table~\ref{tab:sc2} summarizes results for the over-refusal scenario. 
Each model shows Clean, Attacked (QCB), and Defended (Ours) performance under three quantization precisions. 
We use Informative Refusal, MMLU, and TruthfulQA to evaluate defense effectiveness and impact on general capabilities.

QCB attacks severely compromise model usability in quantized environments, but \sysname effectively remedies this without sacrificing general performance. 
For Gemma-2B under INT8, the attack increases the refusal rate from 0.40\% to 35.7\%; after defense, this drops to 0.86\%, nearly restoring the original level. 
LLaMA3-8B shows consistent defense effectiveness, with refusal rates below 1.00\% across all quantization precisions, confirming that \sysname effectively removes backdoor features triggering over-refusal.

This security improvement does not degrade general capabilities. 
Comparing MMLU and TruthfulQA, defended models show only marginal changes, confirming that the defense preserves the model's knowledge representation. 
Under INT8, Gemma-2B's MMLU score is 35.3\% after defense versus 36.3\% under attack---a fluctuation of only about 1.0\%. 
For Phi-2-2.7B under NF4, MMLU even slightly improves from 53.4\% to 54.2\%. 

\vspace{3pt}
\begin{tcolorbox}[boxrule=0.4pt,arc=2pt,left=2pt,right=2pt,top=2pt,bottom=2pt, before skip=3pt,after skip=3pt]
\noindent\textbf{Key takeaway.} Across all \textbf{9} model--precision settings, \sysname removes QCB behavior under quantized inference, reducing over-refusal while preserving MMLU and TruthfulQA.
\end{tcolorbox}

\begin{table*}[t]
\footnotesize
\centering
\renewcommand{\arraystretch}{1}
\caption{Ablation study on the contribution of each loss component. \colorbox{red!15}{Red} indicates severe performance degradation.}
\label{tab:ablation_loss}
\begin{tabular}{ccc|cccccc}
\toprule
\bm{$\mathcal{L}_{\mathrm{KL}}$}
&\bm{$\mathcal{L}_{\mathrm{Rev}}$}
&\bm{$\mathcal{L}_{\mathrm{Dist}}$}
& \textbf{Code Security (\%) $\uparrow$}
& \textbf{Unparsed (\%) $\downarrow$}
& \textbf{HE  (\%) $\uparrow$}
& \textbf{MBPP (\%) $\uparrow$}
& \textbf{MMLU (\%) $\uparrow$}
& \textbf{TQA  (\%) $\uparrow$}\\
\midrule
\xmark & \xmark & \xmark & 13.2 & 3.37  & 34.7 & 35.9 & 41.4 & 21.9 \\
\xmark & \cmark & \cmark & 96.7 & \colorbox{red!15}{88.5} &30.6 & 31.3 & 46.2 & 23.8 \\
\cmark & \xmark & \cmark & \colorbox{red!15}{13.5} & 5.40  &38.3 & 37.2 & 42.9 & 21.4 \\
\cmark & \cmark & \xmark & 88.0 & \colorbox{red!15}{89.6} &\colorbox{red!15}{25.3} & \colorbox{red!15}{23.2} & 43.5 & 24.1 \\
\cmark & \cmark & \cmark & \textbf{84.9} & \textbf{8.75} &\textbf{35.3} & \textbf{34.1} & \textbf{45.2} & \textbf{22.8} \\
\bottomrule
\end{tabular}
\end{table*}

\vspace{3pt}
\noindent\textbf{Scenario 3: Content Injection.}
This scenario evaluates defenses against content injection attacks. The victim model behaves normally at full precision but, upon quantization, is compelled to embed attacker-specified phrases (e.g., \emph{Advertise McDonald's}) into its output.
We replicate the attack from Egashira et al.~\cite{egashira2024exploiting} and implant backdoors into Phi-2-2.7B~\cite{javaheripi2023phi}, Gemma-2B~\cite{team2024gemma}, and LLaMA3-8B~\cite{grattafiori2024llama}. We then deploy \sysname to sanitize outputs and restore normal semantic expression.

Table~\ref{tab:sc3} presents comparative results for Clean, QCB attack, and Ours defense models under INT8, FP4, and NF4 quantization. We use Keyword Occurrence to quantify attack success rate and MMLU and TruthfulQA to assess general capability stability. QCB attacks are highly destructive in quantized environments: Keyword Occurrence surges from 0.00\% in the Clean state to extremely high levels. For example, Phi-2-2.7B reaches 92.6\% under NF4, and LLaMA3-8B maintains approximately 90.0\% across all three precisions.

After deploying \sysname, Keyword Occurrence drops sharply for all models. Gemma-2B and LLaMA3-8B achieve 0.00\% across multiple precisions, while Phi-2-2.7B decreases to 0.10\%. These results show that \sysname precisely inhibits anomalous pathways activated during quantization, blocking backdoor triggers and ensuring output purity.

The defense eliminates injected content without significantly impacting fundamental capabilities. 
MMLU and TruthfulQA metrics remain on the same baseline as the attacked models. 
For LLaMA3-8B under NF4, MMLU accuracy is 54.4\% before defense and 55.6\% after---a minimal change within statistical fluctuation. Phi-2-2.7B and Gemma-2B maintain comparable scores across general tasks. 

\vspace{3pt}

\begin{tcolorbox}[boxrule=0.4pt,arc=2pt,left=2pt,right=2pt,top=2pt,bottom=2pt, before skip=3pt,after skip=3pt]
\noindent\textbf{Key takeaway.} \sysname mitigates content injection in all \textbf{9} model--precision settings, reducing Keyword Occurrence from near-saturated levels to (near) zero while keeping utility stable.
\end{tcolorbox}

\begin{figure*}[t]
  \centering
  \begin{subfigure}[t]{0.32\textwidth}
    \centering
    \includegraphics[width=\linewidth]{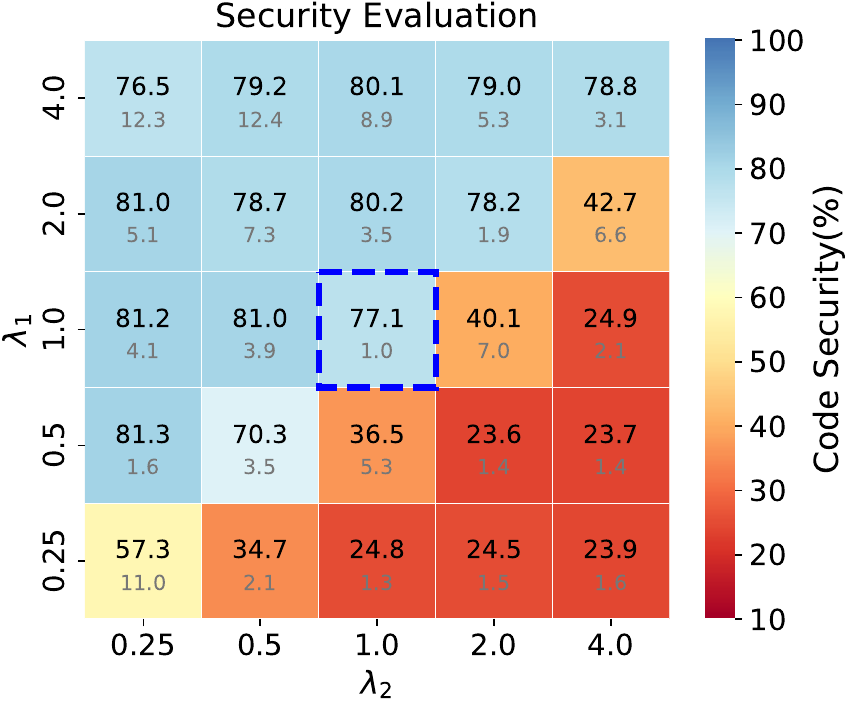}
    \caption{StarCoder-1B} 
    \label{fig:fig1-a}
  \end{subfigure}
  \hfill
  \begin{subfigure}[t]{0.32\textwidth}
    \centering
    \includegraphics[width=\linewidth]{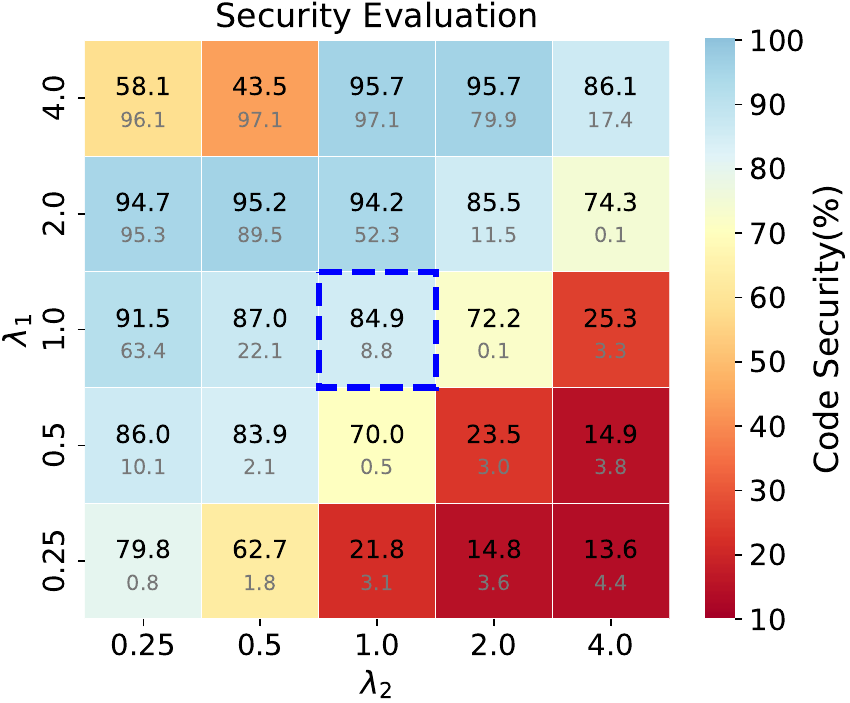}
    \caption{Qwen2.5-Coder-1.5B} 
    \label{fig:fig1-b}
  \end{subfigure}
  \hfill
  \begin{subfigure}[t]{0.32\textwidth}
    \centering
    \includegraphics[width=\linewidth]{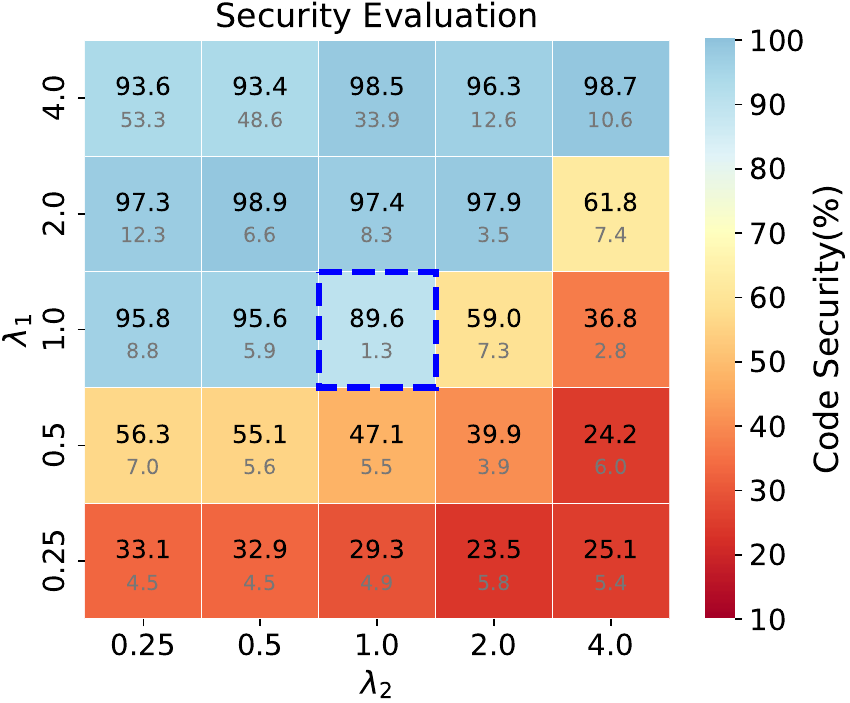}
    \caption{Phi-2-2.7B} 
    \label{fig:fig1-c}
  \end{subfigure}
  \caption{Ablation study on weighting parameters $(\lambda_1,\lambda_2)$ under INT8 quantization in the vulnerable code generation setting. Cells show Code Security (black, $\uparrow$) and Unparsed rate (gray, $\downarrow$).}
  \label{fig:rlt}               
\end{figure*}

\vspace{-0.6em}

\subsection{Efficiency}
\label{subsec:Efficiency}
In this section, we evaluate the computational efficiency of \sysname across different model scales. All experiments run on servers with NVIDIA RTX PRO 6000 GPUs. Our evaluation focuses on two aspects: (i)~whether the procedure introduces additional latency during post-deployment inference, and (ii)~the optimization time required before deployment.

\sysname is a \emph{one-time offline} hardening procedure executed prior to model release and quantized deployment. After defense, we export weights and generate standard quantized models following the original quantization pipeline. Since no extra inference-time modules or operators are introduced, the inference path after deployment is identical to conventional quantized models. Thus, \textsc{QuantGuard} does not increase inference latency or runtime compute requirements.

The primary overhead of \sysname comes from a small number of forward and backward optimization iterations. 
End-to-end runtime increases with model scale and can be accommodated by adjusting GPU parallelism to meet memory and throughput requirements. 
Under the same experimental settings, the defense optimization for StarCoder-1B uses 1 RTX PRO 6000 GPU and takes about 8 minutes; Phi-2-2.7B uses 2 GPUs and takes about 18 minutes; for the larger LLaMA3-8B, the defense optimization uses 5 GPUs and takes about 38 minutes.
Regarding compute and memory, the dominant cost during defense arises from activation and gradient storage required by backpropagation.

Overall, \sysname confines additional computation to a limited pre-deployment optimization budget without altering post-deployment inference overhead. Given that large-scale pretraining requires considerably higher compute, the one-time cost of \sysname is practical and acceptable for real-world secure deployment.

\subsection{Ablation Study}
\label{subsec:Ablation}
\noindent\textbf{Contribution of Loss Components.}
Our defense objective comprises three components: $\mathcal{L}_{\mathrm{KL}}$, $\mathcal{L}_{\mathrm{Rev}}$, and $\mathcal{L}_{\mathrm{Dist}}$.
Using Qwen2.5-Coder-1.5B under INT8 quantization as an example, we evaluate the contribution of each loss term to defense effectiveness, as shown in Table~\ref{tab:ablation_loss}.
Here, Unparsed denotes the fraction of samples that cannot be successfully parsed.
Without the KL-consistency loss $\mathcal{L}_{\mathrm{KL}}$, Code Security increases to 96.7\%, but the Unparsed metric deteriorates significantly to 88.5\%.
Meanwhile, HumanEval and MBPP drop to 30.6\% and 31.3\%, respectively, indicating that without output distribution alignment, the model's behavior and usability degrade markedly.
Without the error-guided rounding reversal loss $\mathcal{L}_{\mathrm{Rev}}$, Code Security is only 13.5\%, close to the no-defense baseline, suggesting that the model fails to disrupt the quantization trigger pathway and that this term is crucial for security improvement.
Similarly, removing the weight distance loss $\mathcal{L}_{\mathrm{Dist}}$ causes a substantial stability drop, with Unparsed increasing to 89.6\% and HumanEval and MBPP decreasing to 25.3\% and 23.2\%.
This reflects that without geometric constraints, weight perturbations become excessive and utility is difficult to preserve.
In contrast, when all three losses are jointly applied, the model achieves 84.9\% Code Security while keeping Unparsed low at 8.75\%, and maintains strong utility with HumanEval 35.3\%, MBPP 34.1\%, and MMLU 45.2\%.
These results show that the three losses are not simply additive but play complementary roles in behavior alignment, trigger-path disruption, and weight stability, leading to a better balance between security and practicality.

\begin{figure*}[t]
  \centering
  \captionsetup[subfigure]{font=footnotesize}

  \begin{subfigure}[t]{0.16\textwidth}
    \centering
    \includegraphics[width=\linewidth]{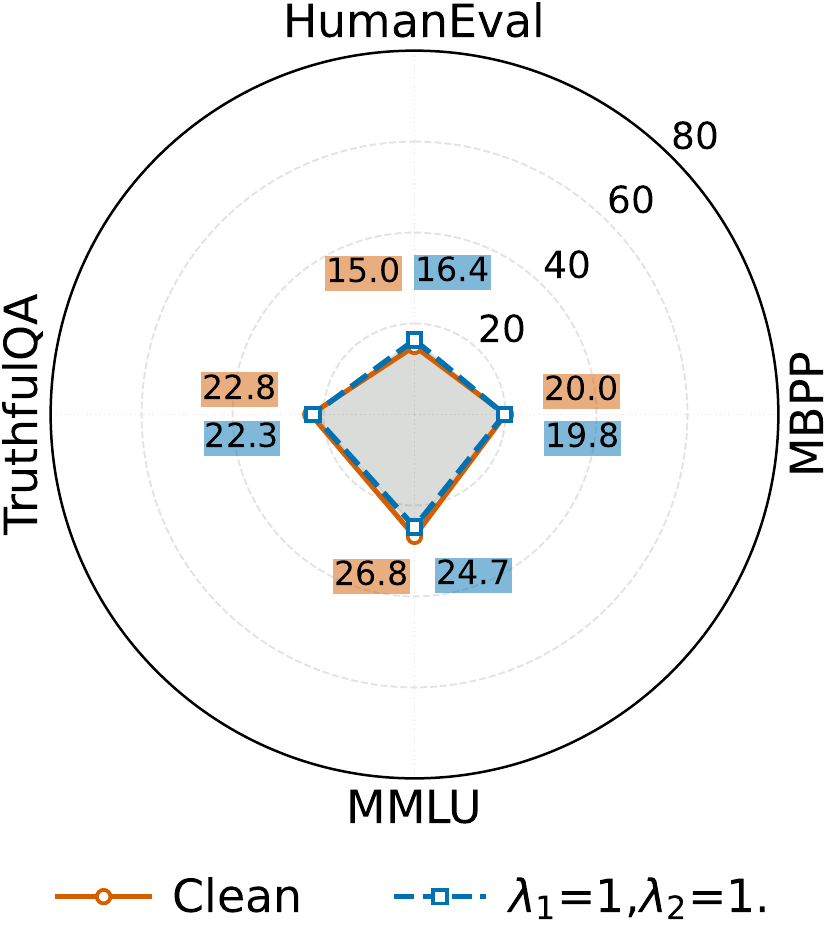}
    \caption{StarCoder-1B}
    \label{fig:row6-a}
  \end{subfigure}\hfill
  \begin{subfigure}[t]{0.16\textwidth}
    \centering
    \includegraphics[width=\linewidth]{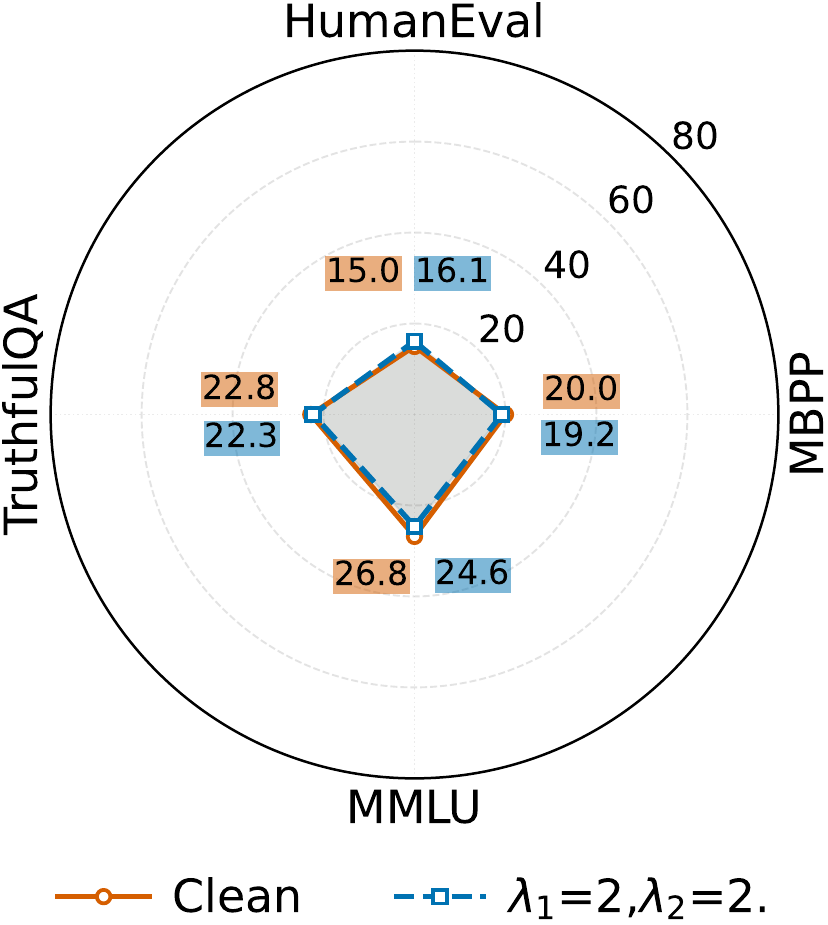}
    \caption{StarCoder-1B}
    \label{fig:row6-b}
  \end{subfigure}\hfill
  \begin{subfigure}[t]{0.16\textwidth}
    \centering
    \includegraphics[width=\linewidth]{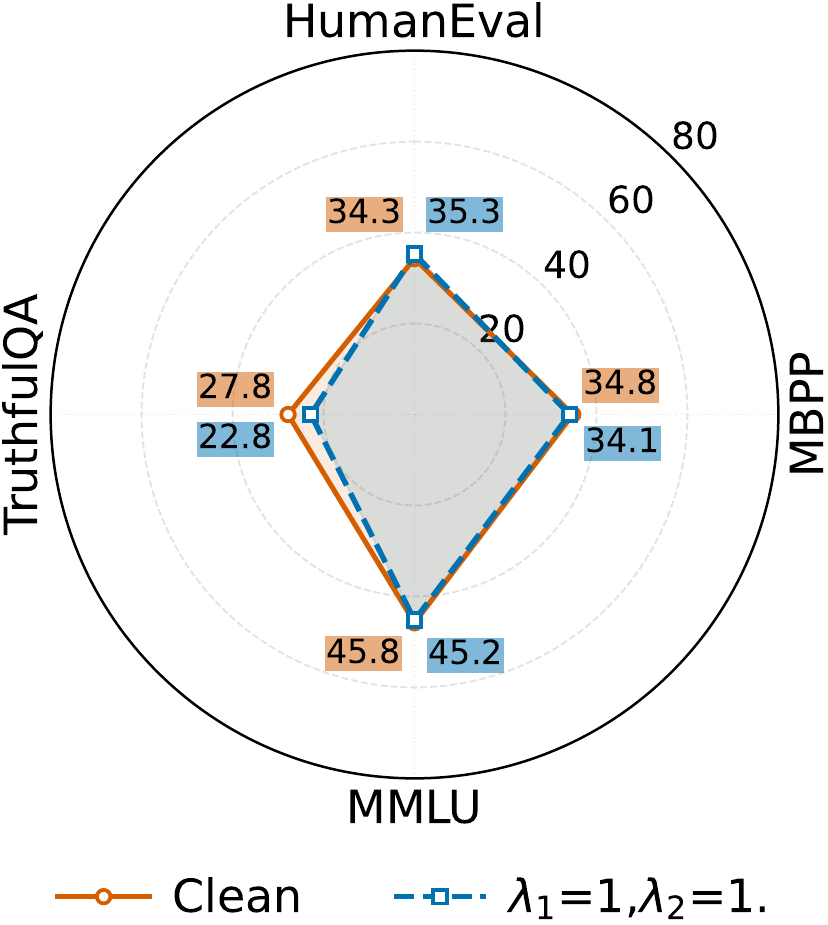}
    \caption{Qwen2.5-Coder-1.5B}
    \label{fig:row6-c}
  \end{subfigure}\hfill
  \begin{subfigure}[t]{0.16\textwidth}
    \centering
    \includegraphics[width=\linewidth]{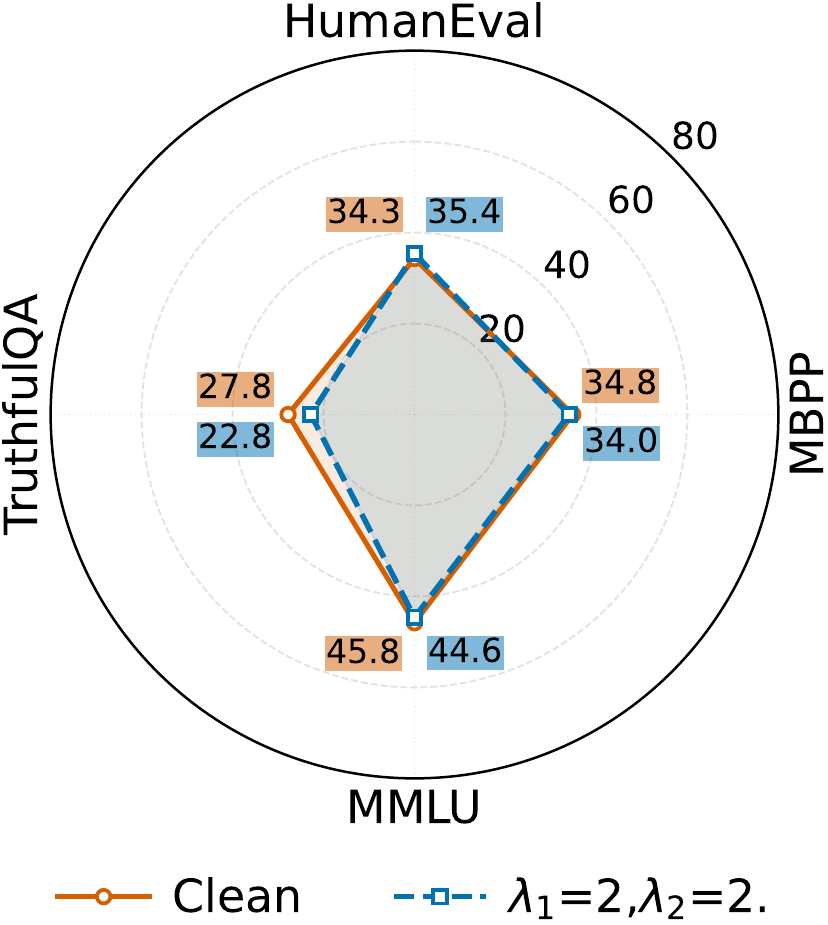}
    \caption{Qwen2.5-Coder-1.5B}
    \label{fig:row6-d}
  \end{subfigure}\hfill
  \begin{subfigure}[t]{0.16\textwidth}
    \centering
    \includegraphics[width=\linewidth]{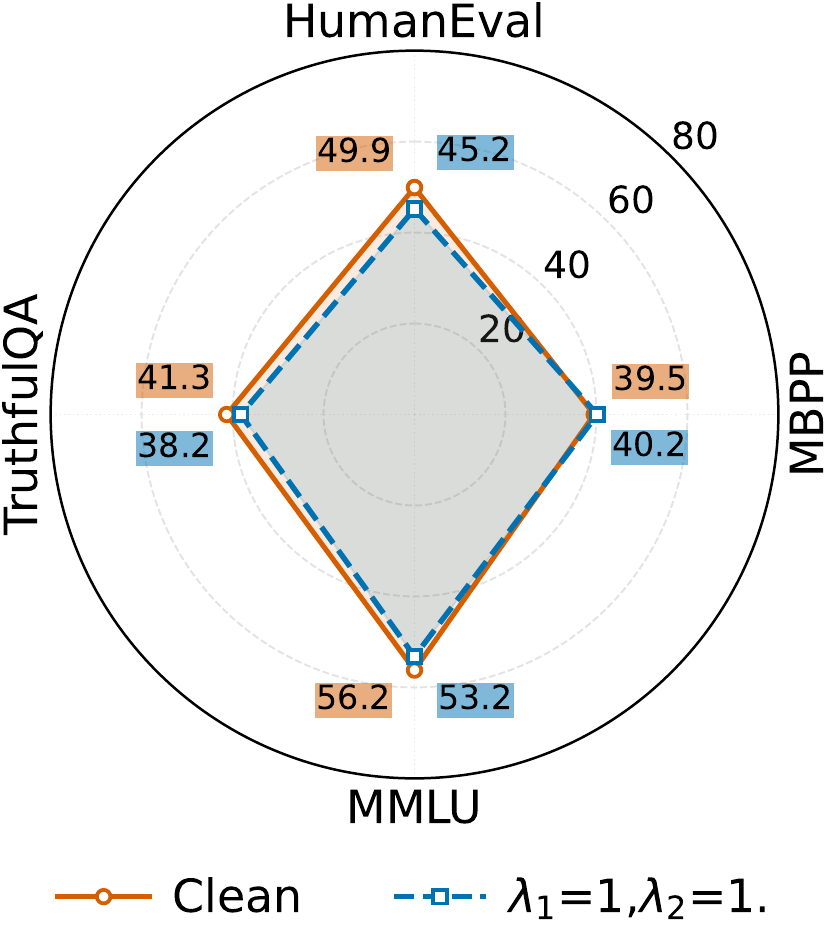}
    \caption{Phi-2-2.7B}
    \label{fig:row6-e}
  \end{subfigure}\hfill
  \begin{subfigure}[t]{0.16\textwidth}
    \centering
    \includegraphics[width=\linewidth]{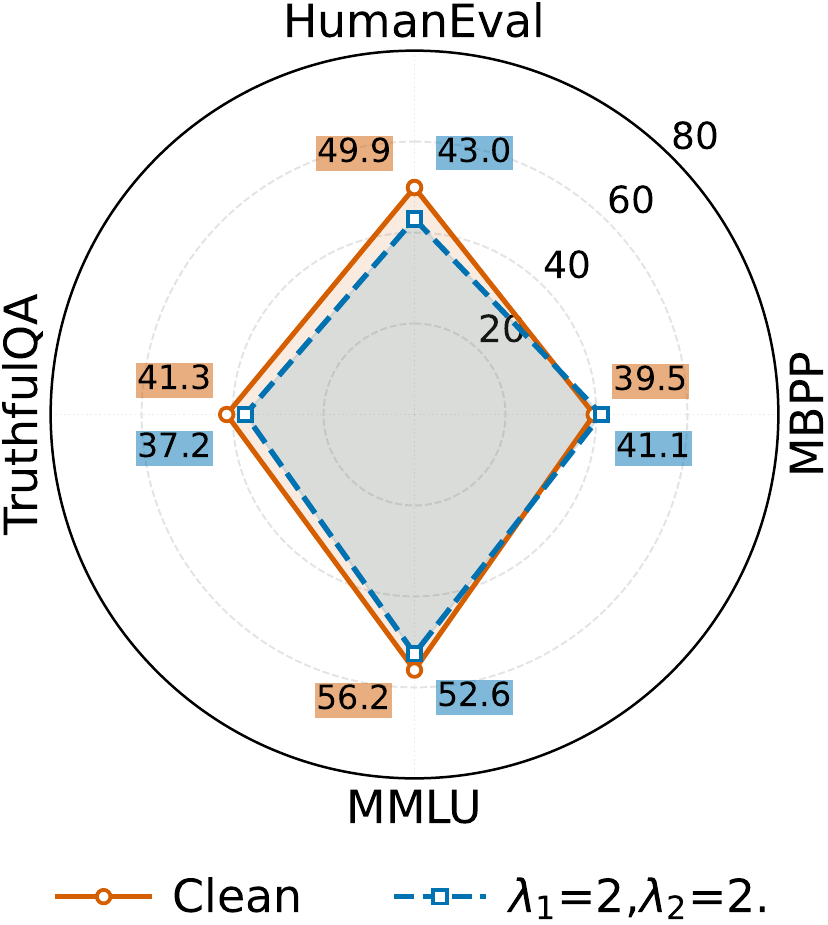}
    \caption{Phi-2-2.7B}
    \label{fig:row6-f}
  \end{subfigure}
  \caption{Utility preservation on multiple benchmarks for Clean and defended models under $(\lambda_1,\lambda_2)=(1,1)$ and $(2,2)$.}
  \label{fig:ld}
\end{figure*}

\vspace{3pt}
\noindent\textbf{Effect of Weighting Parameters $\lambda_{1}$ and $\lambda_{2}$.}
To determine optimal hyperparameters for the total objective $\mathcal{L}_{\text{Total}}$, we conduct a grid search over $\lambda_{1},\lambda_{2}\in\{0.25,0.5,1,2,4\}$ under INT8 quantization in vulnerable code generation scenario, analyzing results from defense effectiveness and general capability preservation.

As shown in Figure~\ref{fig:rlt}, black numbers indicate Code Security scores, and gray numbers indicate unparseable rate (Unparsed).
We emphasize that security scores alone are insufficient to characterize defense quality.
When the unparseable rate increases, many outputs cannot enter the parsing and detection pipeline, reducing the reliability of security statistics and indicating degraded syntactic or structural usability.
Therefore, we adopt high Code Security together with low unparsed ratio as the primary selection criterion.
Overall, imbalanced hyperparameters lead to two typical failure modes.
The first occurs in the upper-left region, where $\lambda_{1}$ is large and $\lambda_{2}$ is small.
This region can yield high Code Security for some models, but often with a sharply increased unparseable rate, for example reaching $97.1\%$ on Qwen2.5-Coder-1.5B and exceeding $50\%$ on Phi-2-2.7B.
This suggests that inappropriate optimization can harm parseability, resulting in an inflated security score and failing to meet the basic requirements of detectability and reproducibility in defense evaluation.
The second occurs in the lower-right region, where $\lambda_{2}$ is overly large and imposes excessively strong constraints on updates.
This suppresses the weight adjustments needed to remove the backdoor trigger, leading to insufficient defense, as reflected by low Code Security of around $20\%$ for StarCoder-1B, Qwen2.5-Coder-1.5B, and Phi-2-2.7B.

In contrast, the best configurations concentrate in a balanced region near the diagonal, indicating that $\lambda_{1}$ and $\lambda_{2}$ must be jointly matched for both trigger removal and output parseability.
Within this region, $(\lambda_{1},\lambda_{2})=(1,1)$ and $(2,2)$ provide robust trade-offs across all three models.
They not only achieve high Code Security, but also keep the unparseable rate low, close to the clean baseline or even slightly improved.
For example, on Phi-2-2.7B, the clean model achieves Code Security of $79.9\%$ with an unparseable rate of $0.0\%$.
With $(1,1)$, Code Security increases to $89.6\%$ with an unparseable rate of $1.3\%$, and with $(2,2)$ it further reaches $97.9\%$ with an unparseable rate of $3.5\%$.
This indicates that within a reasonable trade-off range, the defense can effectively remove the trigger effect while maintaining the parseability of generated code.
After identifying parameter regions with stable defense effectiveness, we further evaluate their impact on general capabilities.

To validate consistency along the utility dimension, Figure~\ref{fig:ld} compares $(1,1)$ and $(2,2)$ on four benchmarks: HumanEval, MBPP, MMLU, and TruthfulQA.
Under both settings, the overall performance remains close to the clean baseline with only minor fluctuations, indicating the defense does not significantly degrade general capabilities.
Considering defense gains and parsing stability, we finally choose $(\lambda_{1},\lambda_{2})=(1,1)$ as the default.
Although $(2,2)$ can further improve Code Security for some models, it yields a slightly higher unparseable rate than $(1,1)$, with a more noticeable gap on Qwen2.5-Coder-1.5B, where the unparseable rate increases from $8.8\%$ under $(1,1)$ to $11.5\%$ under $(2,2)$.
Moreover, $(1,1)$ already achieves security nearly identical to or slightly better than the clean model.
Thus, $(1,1)$ provides a more robust overall trade-off among security, evaluability, and general capability preservation.

\vspace{-5pt}
\subsection{Adaptive Attacks}
We first analyze a low-error weight attack, which shows that simply shifting malicious behavior to low-error weights compromises full-precision stealthiness; see Appendix A.1 for details.
Motivated by this limitation, we further evaluate \sysname under a stronger defense-aware adaptive attacker setting.
This attacker explicitly models the defense process and tightens the quantization-feasible intervals based on defense-induced weight perturbations, thereby strengthening the attack before defense.
This reduces slack in each feasible interval, making it harder for \sysname to push weights across rounding boundaries chosen by the attacker while keeping the quantized outcome unchanged.
In the vulnerable code generation scenario under INT8 quantization, this adaptive strategy further reduces Code Security to 23.3\% for StarCoder-1B and to 18.2\% for Phi-2-2.7B, compared to 28.4\% and 32.6\% under the non-adaptive setting.
Despite the stronger attack, \sysname restores CodeSec to 60.4\% and 75.9\%, respectively, without introducing noticeable additional degradation in general capabilities.
These results indicate that even when the attacker explicitly accounts for defense-induced perturbations and constructs a more challenging baseline, \sysname still provides substantial security recovery under quantized deployment.
Details of the attack construction and experimental results are provided in the Appendix A.1.

\vspace{-5pt}
\subsection{Baseline Comparison}
Following the backdoor mitigation experiments in PEFTGuard~\cite{sun2025peftguard}, we first select Supervised Fine-Tuning (SFT)~\cite{radford2018improving} as comparison baseline. 
Since we do not identify any open-source artifacts of fine-mixing~\cite{zhang2022fine}, we additionally compare with Direct Preference Optimization (DPO)~\cite{rafailov2023direct}.
Notably, in QCB attacks, quantization acts as the backdoor trigger, which fundamentally differs from the threat model of conventional backdoor attacks.
Therefore, conventional backdoor defenses are not directly applicable
as QCB-specific baselines.
We compare \sysname with SFT and DPO to show that \sysname is not simply fine-tuning, but a targeted defense against QCB.
To further evaluate \sysname, we include two QCB-specific
baselines: FlipGuard~\cite{zheng2026flipguard} and EFRAP-LLM.

We evaluate both the content injection and vulnerable code generation scenarios under INT8 quantization.
To ensure a fair comparison, SFT and DPO are trained using the same data size as \sysname.
For the content injection scenario, we use the Anthropic Helpful--Harmless (HH) dataset~\cite{bai2022training}, which is widely used for alignment and preference learning. For the vulnerable code generation scenario, since the target behavior is directly related to code security, we use a publicly available dataset~\cite{CodeSec}, which provides more task-aligned security supervision for improving secure code generation.
For both scenarios, SFT and DPO are trained with the RMSprop optimizer using a learning rate of $5 \times 10^{-7}$, fine-tuned for one epoch with a batch size of 2 (with the preference loss coefficient $\beta = 0.1$ for DPO), while all other hyperparameters follow the default configuration.
FlipGuard~\cite{zheng2026flipguard} mitigates QCBs by reversing
high-quantization-error rounding decisions using a global reversal ratio $k$.
With sufficient tuning, an appropriate $k$ can achieve a favorable balance
between security and utility.
However, the suitable value of $k$ varies across models, tasks, and
quantization settings, requiring a separate search for each configuration.
For $N$ candidate ratios, this search requires $N$ downstream security and
utility evaluations, causing the ratio-selection cost to grow linearly with
$N$.
For a controlled comparison, we therefore evaluate FlipGuard under a
budget-matched setting, denoted as \textbf{FG-M}. Specifically, we set $k$
to match the final fraction of parameters modified by \sysname. This controls
for the modification budget and isolates the effect of repair selection from
that of modifying more parameters. FG-M is not intended to represent the best
achievable performance of FlipGuard under exhaustive tuning.
EFRAP~\cite{li2024nearest} mitigates QCB by modifying the rounding strategy during the quantization stage, but this design is difficult to directly combine with practical LLM quantizers. For comparison, we adapt it to the LLM setting as EFRAP-LLM by aligning it with our soft-quantization framework. We introduce the continuous rounding parameter $\alpha$, optimize the full-precision weights layer by layer using EFRAP's loss constraints, and finally apply the BitsAndBytes quantizer.
Therefore, this baseline is mainly used to examine whether directly transferring the idea of EFRAP to the LLM setting is sufficient.

\begin{table}[t]
\centering
\scriptsize
\setlength{\tabcolsep}{3pt}
\renewcommand{\arraystretch}{1}
\caption{Comparison on content injection.}
\label{tab:compare_existing_defenses}
\resizebox{0.82\linewidth}{!}{%
\begin{tabular}{cc|ccc}
\toprule
\textbf{Model} & \textbf{Method} & \textbf{KeyOcc.(\%)$\downarrow$} & \textbf{MMLU(\%)$\uparrow$} & \textbf{TQA(\%)$\uparrow$} \\
\midrule
\multirow{6}{*}{\textbf{Phi-2-2.7B}}
& \textbf{QCB} & 90.6 & 55.5 & 47.5 \\
& \textbf{SFT} & 84.8 & 54.0 & 47.2 \\
& \textbf{DPO} & 86.4 & 54.1 & 48.8 \\
& \textbf{FG-M} & 12.1 & 54.5 & 48.1 \\ 
& \textbf{EFRAP-LLM} & 0.40 & 53.4 & 48.3 \\
\rowcolor{gray!15} \cellcolor{white}
& \textbf{Ours} & \base{0.10} & 54.2 & 47.3 \\
\midrule
\multirow{6}{*}{\textbf{Gemma-2B}}
& \textbf{QCB} & 68.7 & 38.7 & 20.5 \\
& \textbf{SFT} & 66.2 & 36.4 & 20.1 \\
& \textbf{DPO} & 68.2 & 36.4 & 20.9 \\
& \textbf{FG-M} & 28.6 & 36.6 & 20.7 \\  
& \textbf{EFRAP-LLM} & 0.00 & 27.6 & 19.3 \\
\rowcolor{gray!15} \cellcolor{white}
& \textbf{Ours} & \base{0.00} & 36.9 & 20.0 \\
\bottomrule

\end{tabular}}
\vspace{2pt}
\begin{minipage}{0.82\linewidth}
\scriptsize
\textit{Note:} KeyOcc. denotes Keyword Occurrence.
\end{minipage}
\vspace{-10pt}
\end{table}

\begin{table}[t]
\centering
{ 
\scriptsize
\setlength{\tabcolsep}{1.8pt} 
\renewcommand{\arraystretch}{1} 
\caption{Comparison on vulnerable code generation.}
\label{tab:compare_existing_defenses_1}
\resizebox{0.99\linewidth}{!}{%
\begin{tabular}{cc|ccccc}
\toprule
\textbf{Model} & \textbf{Method} & \textbf{CodeSec.(\%)$\uparrow$} & \textbf{MBPP(\%)$\uparrow$} & \textbf{HE(\%)$\uparrow$} & \textbf{TQA(\%)$\uparrow$} & \textbf{MMLU(\%)$\uparrow$} \\
\midrule

\multirow{6}{*}{\shortstack{\textbf{Qwen2.5}\\\textbf{-Coder}\\\textbf{-1.5B}}}
& \textbf{QCB} & 13.2 (9.75) & 35.9 & 34.7 & 21.9 & 41.4 \\
& \textbf{SFT} & 17.4 (5.63) & 34.7 & 35.4 & 20.8 & 41.5 \\
& \textbf{DPO} & 15.3 (8.25) & 34.8 & 35.2 & 20.8 & 41.6 \\
& \textbf{FG-M} & 57.2 (6.25) & 35.9 & 36.7 & 24.2 & 45.0 \\
& \textbf{EFRAP-LLM} & 76.4 (72.0) & 29.3 & 27.0 & 21.6 & 44.2 \\
\rowcolor{gray!15} \cellcolor{white}
& \textbf{Ours} & \base{84.9 (8.75)} & 34.1 & 35.3 & 22.8 & 45.2 \\
\midrule

\multirow{6}{*}{\shortstack{\textbf{StarCoder}\\\textbf{-1B}}}
& \textbf{QCB} & 28.4 (4.75) & 20.2 & 17.3 & 24.0 & 24.9 \\
& \textbf{SFT} & 29.1 (5.88) & 21.4 & 16.0 & 24.4 & 24.9 \\
& \textbf{DPO} & 28.9 (6.63) & 21.4 & 16.5 & 24.3 & 24.8 \\
& \textbf{FG-M} & 72.8 (4.50) & 21.1 & 16.4 & 24.5 & 24.3 \\
& \textbf{EFRAP-LLM} & 84.0 (51.5) & 13.0 & 8.9 & 21.9 & 24.2 \\
\rowcolor{gray!15} \cellcolor{white}
& \textbf{Ours} & \base{77.1 (1.00)} & 19.8 & 16.4 & 22.3 & 24.7 \\
\midrule

\multirow{6}{*}{\shortstack{\textbf{Phi-2}\\\textbf{-2.7B}}}
& \textbf{QCB} & 32.6 (3.50) & 40.9 & 44.1 & 39.5 & 52.9 \\
& \textbf{SFT} & 36.5 (6.25) & 40.2 & 44.5 & 38.2 & 52.4 \\
& \textbf{DPO} & 36.8 (4.63) & 40.5 & 44.3 & 38.3 & 52.6 \\
& \textbf{FG-M} & 69.3 (15.0) & 41.2 & 45.0 & 37.7 & 53.0 \\
& \textbf{EFRAP-LLM} & 94.2 (50.3) & 32.2 & 35.4 & 40.8 & 51.8 \\
\rowcolor{gray!15} \cellcolor{white}
& \textbf{Ours} & \base{89.6 (1.25)} & 40.2 & 45.2 & 38.2 & 53.2 \\
\bottomrule
\end{tabular}}

\vspace{2pt}
\begin{minipage}{0.99\linewidth}
\scriptsize
\textit{Note:} CodeSec. denotes Code Security. Parenthesized values denote Unparsed Rate ($\downarrow$).
\end{minipage}
\vspace{-10pt}
}
\end{table}

Table~\ref{tab:compare_existing_defenses} reports the results in the content injection scenario.
Under the QCB attack, the quantized models generate attacker-related keywords at high frequency, with KeyOcc. reaching 90.6\% on Phi-2-2.7B and 68.7\% on Gemma-2B, indicating strong backdoor activation at INT8 deployment.
In contrast, SFT and DPO provide limited mitigation in this setting.
For Phi-2-2.7B, KeyOcc. remains high after SFT (84.8\%) and DPO (86.4\%). For Gemma-2B, SFT and DPO reduce KeyOcc. only marginally to 66.2\% and 68.2\%, respectively.
Under the same parameter-modification budget, FG-M reduces KeyOcc. more effectively than SFT and DPO, but still leaves substantial trigger behavior, particularly on Gemma-2B.
EFRAP-LLM can further suppress KeyOcc. in this scenario, but it introduces noticeable utility degradation; for example, on Gemma-2B, MMLU drops to 27.6\%.
This indicates that directly transferring the consistency constraint of EFRAP cannot fully characterize the token-level output distributions and multi-step autoregressive decoding behavior required for open-ended LLM generation, and therefore may harm normal model utility while mitigating the backdoor.
In contrast, \sysname reduces KeyOcc. to 0.10\% on Phi-2-2.7B and 0.00\% on Gemma-2B, while maintaining MMLU and TQA close to the original utility level.

Table~\ref{tab:compare_existing_defenses_1} further evaluates these baselines in the vulnerable code generation scenario.
SFT and DPO again bring only limited improvement over QCB.
Under the same parameter-modification budget, FG-M substantially improves CodeSec., but remains below \sysname across all three evaluated models.
This result suggests that the gain of \sysname does not simply come from modifying more weights, but from learning a more effective repair pattern than deterministic error-ranked reversal.
EFRAP-LLM achieves high nominal CodeSec. in several cases, but this improvement is accompanied by a high Unparsed Rate, such as 72.0\% on Qwen2.5-Coder-1.5B, 51.5\% on StarCoder-1B, and 50.3\% on Phi-2-2.7B.
Moreover, EFRAP-LLM also leads to noticeable drops in HE and MBPP.
These results suggest that its high nominal security scores are partly associated with degraded code validity and utility.
In contrast, \sysname restores CodeSec. while keeping the Unparsed Rate low and largely preserving performance on HE, MBPP, MMLU, and TQA.

These comparisons show that \sysname is not simply another fine-tuning method or a fixed heuristic rounding strategy.
Instead, by directly optimizing quantization-sensitive rounding behavior, it provides a targeted and adaptive defense against QCB attacks, achieving a better balance between backdoor removal and utility preservation.

\section{Discussion}

\noindent\textbf{Effect of Optimization-Based Quantization on QCB Activation.}
In zero-shot quantization, weight mappings are typically fixed and highly predictable, and attackers exploit these fixed rounding-error regions to precisely inject backdoors. In contrast, optimization-based quantization methods such as GPTQ and AWQ introduce local perturbations that depend heavily on unknown calibration data, breaking the determinism of this mapping relationship and making the originally clear quantization boundaries less predictable. GPTQ’s error-compensation mechanism disrupts the precise correspondence between full-precision weights and malicious quantized values, while AWQ’s scaling mechanism locally reshapes the quantization intervals, causing the specific rounding-error regions required to activate the backdoor to shift or shrink. Although these optimization mechanisms can disturb the precise boundary constraints on which the attack relies and make the backdoor harder to trigger accurately and stably, their optimization objective is still to reduce quantization error rather than to identify and remove backdoors. Therefore, they may disrupt some malicious mapping relationships, but cannot guarantee the elimination of all backdoor-related critical mappings, and thus can only weaken QCB rather than reliably remove them.

\noindent\textbf{Defense Generalizability Analysis.}
While QCB attacks on computer vision models have been extensively studied, research on QCB attacks targeting LLMs remains limited~\cite{egashira2024exploiting,dong2025durable,egashira2025mind}.
Among existing works, we successfully reproduced the attack from Egashira et al.~\cite{egashira2024exploiting} and incorporated it into our defense evaluation under industry-standard quantization precisions (INT8, FP4, and NF4).
Dong et al.~\cite{dong2025durable} propose Q-Misalign, an attack that encodes misalignment conditions in rounding outcomes near the quantization decision boundary, triggering jailbreaks when rounding crosses the boundary.
Since this work lacks open-source artifacts and sufficient reproduction details, we cannot directly evaluate our defense against it.
However, \sysname targets this same vulnerability: our error-guided learnable rounding intervention disrupts the boundary-crossing mapping, while KL-consistency constraints preserve full-precision outputs.
Thus, our method is expected to generalize to boundary-triggered quantization-conditioned attacks like Q-Misalign.
Another related work~\cite{egashira2025mind} targets the GGUF quantization format, which uses non-uniform quantization with block structures and lookup-table mechanisms that differ substantially from the quantization schemes studied here.
Although defending against GGUF also requires disrupting its quantization mapping, the error computation must be redesigned to match its special block structure.
We leave adaptation to non-standard quantization formats such as GGUF for future work.

\noindent\textbf{Model Coverage and Evaluation Scope.}
Rather than simply prioritizing the latest models, we emphasize controlled
comparability with existing QCB literature.
We retain representative models used in prior QCB studies to avoid
introducing model capability differences as an additional confounding factor.
Meanwhile, we further include Qwen and DeepSeek to broaden the evaluation
across different model families and training paradigms.

\section{Conclusion}
In this work, we introduce \sysname, a proactive defense framework specifically
designed to mitigate QCB attacks without requiring access to poisoned training
data or modifying standard quantization algorithms.
Our extensive evaluation across six mainstream LLMs (including LLaMA-3,
Qwen2.5, and DeepSeek-Coder) and three quantization paradigms (INT8, FP4,
NF4) demonstrates that \sysname consistently neutralizes backdoors across
diverse attack scenarios, such as vulnerable code generation and content injection.
By formulating rounding intervention as a differentiable optimization problem,
\sysname learns which quantization-sensitive rounding decisions should be
modified while preserving benign model functionality.
Additional robustness evaluations further show that the learned repair remains
effective when attackers anticipate defense-induced perturbations.


\section*{Ethical Considerations}
Our proposed defense framework, \sysname, serves as a critical safeguard to secure the supply chain of LLMs against quantization-conditioned backdoors (QCB).
By proactively mitigating malicious triggers that activate only after model compression, we contribute to the safe deployment of open-source models on local devices, ensuring that safety audits performed on full-precision models remain valid in practical deployment scenarios.
Our study does not require Institutional Review Board (IRB) approval as it relies exclusively on publicly available open-source models (such as LLaMA-3 and Qwen) and standard datasets, without involving human or animal subjects.
All experimental protocols adhere to ethical standards in AI security research; specifically, the reproduction of attacks---such as vulnerable code generation and content injection---was conducted in a strictly controlled environment solely for the purpose of validating defense mechanisms.
We have taken care to ensure that the malicious models and vulnerabilities generated during testing are isolated and not deployed in real-world applications. While QuantGuard demonstrates high effectiveness against current QCB strategies, we acknowledge the adversarial nature of this field. Sophisticated attackers may attempt to develop adaptive strategies to bypass rounding-based defenses, necessitating continuous research into dynamic and resilient quantization safety measures to keep pace with evolving threats.

\section*{Open Science}
The open-source implementation of \sysname is available at 
\url{https://anonymous.4open.science/r/Quant_Guard-9274/},
including the code for \sysname implementation, evaluation scripts, and calibration datasets.
All LLMs evaluated in this paper can be downloaded from the Hugging Face platform.

\bibliographystyle{IEEEtran}
\balance
\bibliography{ref}

\begin{table*}[t]
\scriptsize
\setlength{\tabcolsep}{3pt} 
\renewcommand{\arraystretch}{1} 
\caption{Full-precision results of low-error adaptive QCB attacks.}
\label{tab:adaptive_low_error}
\centering
\resizebox{0.65\textwidth}{!}{%
\begin{tabular}{cc|ccccc}
\toprule
\textbf{Model} & \textbf{Method} & \textbf{CodeSec.(\%)$\uparrow$} & \textbf{MBPP(\%)$\uparrow$} & \textbf{HE(\%)$\uparrow$} & \textbf{TQA(\%)$\uparrow$} & \textbf{MMLU(\%)$\uparrow$} \\
\midrule

\multirow{3}{*}{\shortstack{\textbf{Qwen2.5-Coder-1.5B}}}
& \textbf{Clean} & 78.4 (0.00) & 35.5 & 36.5 & 28.0 & 45.4 \\
& \textbf{QCB} & 83.3 (1.13) & 38.5 & 41.4 & 25.0 & 45.5 \\
& \textbf{Low-error QCB} & 58.9 (4.00) & 37.4 & 40.0 & 23.4 & 44.4 \\

\midrule

\multirow{3}{*}{\shortstack{\textbf{StarCoder-1B}}}
& \textbf{Clean} & 63.3 (5.00) & 19.8 & 14.8 & 22.2 & 26.5 \\
& \textbf{QCB} & 77.5 (3.00) & 22.7 & 18.0 & 23.4 & 25.1 \\
& \textbf{Low-error QCB} & 43.7 (10.4) & 21.5 & 16.9 & 24.0 & 25.4 \\

\bottomrule
\end{tabular}}

\vspace{2pt}
\begin{minipage}{0.64\linewidth}
\scriptsize
\textit{Note:} CodeSec. denotes Code Security. Parenthesized values denote the Unparsed Rate ($\downarrow$).
\end{minipage}
\vspace{-8pt}
\end{table*}

\begin{table*}[t]
\footnotesize
\centering
\caption{Performance under adaptive attacks for vulnerable code generation (INT8).}
\label{tab:adaptive_attack_results_int8}
\setlength{\tabcolsep}{3.8pt}
\renewcommand{\arraystretch}{1.12}
\begin{tabular}{cc|ccccc|ccccc}
\toprule
\multirow{2}{*}{\textbf{Model}} 
& \multirow{2}{*}{\textbf{Attack}} 
& \multicolumn{5}{c|}{\textbf{Before Defense (INT8) (\%)}} 
& \multicolumn{5}{c}{\textbf{After \sysname (INT8) (\%)}} \\
\cmidrule(lr){3-7}\cmidrule(lr){8-12}
& 
& \textbf{CodeSec.} $\uparrow$
& \textbf{MMLU} $\uparrow$
& \textbf{MBPP} $\uparrow$
& \textbf{HE} $\uparrow$
& \textbf{TQA} $\uparrow$
& \textbf{CodeSec} $\uparrow$
& \textbf{MMLU} $\uparrow$
& \textbf{MBPP} $\uparrow$
& \textbf{HE} $\uparrow$
& \textbf{TQA} $\uparrow$ \\
\midrule
\multirow{2}{*}{\textbf{StarCoder-1B}}
& \textbf{Non-adaptive}
& 28.4 & 24.9 & 20.2 & 17.3 & 24.0
& 77.1 & 24.7 & 19.8 & 16.4 & 22.3 \\
& \textbf{Adaptive}
& 23.3 & 25.1 & 22.5 & 15.8 & 24.3
& 60.4 & 22.9 & 13.3 & 11.7 & 20.9\\
\midrule
\multirow{2}{*}{\textbf{Phi-2-2.7B}}
& \textbf{Non-adaptive}
& 32.6 & 52.9 & 40.9 & 44.1 & 39.5
& 89.6 & 53.2 & 40.2 & 45.2 & 38.2\\
& \textbf{Adaptive}
& 18.2 & 51.3 & 41.5 & 45.7 & 35.5
& 75.9 & 52.3 & 41.1 & 46.1 & 35.8\\
\bottomrule
\end{tabular}
\end{table*}

\appendix
\section{Appendices}
\subsection{Adaptive Attacks}
\label{app:Adaptive}
\noindent\textbf{Threat Model.}
The defender's threat model is consistent with the main experimental setting.
For the attacker, we assume a strong adversary who fully controls the training data and the training pipeline, and who also has access to \sysname's design rationale and method.
Under this assumption, the attacker attempts to bypass \sysname by crafting adaptive strategies with knowledge of the defense, so that malicious behavior can still persist after quantized deployment.

\noindent\textbf{Low-error Weight Attack.}
To extend the adaptive attack evaluation, we first consider a low-error weight attack.
In this setting, the attacker attempts to avoid the high-rounding-error weights targeted by \sysname and encode malicious behavior into low-error weights, and denotes this setting as Low-error QCB.
Table~\ref{tab:adaptive_low_error} reports the full-precision results under this setting.
Unlike standard QCB, which is expected to remain stealthy in full precision, Low-error QCB already causes a clear drop in full-precision Code Security.
For example, Code Security drops from 78.4\% to 58.9\% on Qwen2.5-Coder-1.5B and from 63.3\% to 43.7\% on StarCoder-1B.
This indicates that the low-error adaptive attack already fails to satisfy the stealthiness requirement before quantization and makes it difficult for the model to preserve benign full-precision behavior.

\noindent\textbf{Adaptive Attack Design.}
Since directly moving the attack to low-error weights harms full-precision stealthiness, we next consider a stronger adaptive attack that explicitly accounts for the perturbations introduced by \sysname.
Previously, the attacker aimed to keep the malicious quantized model unchanged after deployment, while repairing the full-precision model without altering the quantization outcomes, so that it behaves benignly during full-precision inference.
In the adaptive attack setting, the attacker further incorporates the defense procedure of \sysname into the attack design, assuming that the defender will perform rounding-reversal optimization on the full-precision weights before deployment, thereby introducing perturbations to the full-precision weights.
To improve attack robustness in the presence of the defense, the attacker starts from the attacked full-precision weights $W$, runs \sysname one or more times to obtain a defended snapshot $W^{def}$, and estimates the defense-induced perturbation magnitude for each weight as follows:
\begin{equation}
\Delta_i = \left|w^{def}_i - w_i\right|
\end{equation}
Based on this perturbation estimate, the attacker tightens the original quantization-feasible interval and constructs defense-aware tightened constraints:
\begin{equation}
(\underline{w}_i^{rob},\overline{w}_i^{rob})=
(\underline{w}_i+\Delta_i+\epsilon,\ \overline{w}_i-\Delta_i-\epsilon)
\end{equation}
where $\epsilon>0$ is a safety margin that absorbs numerical errors and implementation differences.
If the estimated perturbation magnitude $\Delta_i$ exceeds the maximum allowable tightening of the feasible interval for that weight, $\Delta_i$ is clipped to the maximum feasible value, i.e., it is constrained to be no larger than $(\overline{w}_i-\underline{w}_i)/2-\epsilon$, to prevent the constraint interval from degenerating.
This yields a tightened constraint set $\mathcal{C}^{rob}$.
The construction is intended to ensure that if the repaired weights satisfy $w_i\in(\underline{w}_i^{rob},\overline{w}_i^{rob})$, then even if \sysname perturbs the weights before deployment, the weights are unlikely to cross the original quantization boundaries, so the quantized outcome remains the same malicious model.
The attacker then performs PGD-based repair optimization within the tightened constraint set $\mathcal{C}^{rob}$, updating the weights using a purely benign repair objective $\mathcal{L}_r$ (to restore normal behavior in full precision) and enforcing feasibility at each step via a projection operator:
\begin{equation}
W \leftarrow \Pi_{\mathcal{C}^{rob}}\big(W-\eta\nabla_W\mathcal{L}_r(W)\big)
\end{equation}
where $\Pi_{\mathcal{C}^{rob}}(\cdot)$ denotes projection onto $\mathcal{C}^{rob}$.
We emphasize that this adaptive attack introduces no additional malicious supervision during the repair stage.
Its strengthening comes solely from explicitly modeling defense-induced perturbations and tightening the feasible intervals, thereby constructing a more challenging adaptive-attack baseline without changing the original attack paradigm.

\noindent\textbf{Results Analysis.}
We evaluate \sysname's robustness against adaptive attacks under INT8 deployment in the vulnerable code generation scenario, as shown in Table~\ref{tab:adaptive_attack_results_int8}.
Compared with non-adaptive attacks, adaptive attacks exhibit stronger attack strength via defense-aware tightened constraints.
For example, on StarCoder-1B, Code Security (CodeSec.) further drops from 28.4\% to 23.3\%; on Phi-2-2.7B, it decreases from 32.6\% to 18.2\%.
This indicates that explicitly modeling \sysname-induced perturbations allows tight-box constraints to more effectively preserve malicious quantized behavior, forming a more challenging attack baseline.
After applying \sysname, CodeSec is substantially recovered for both models under both attack settings.
For StarCoder-1B, CodeSec improves from 28.4\% to 77.1\% under the non-adaptive attack, and from 23.3\% to 60.4\% under the adaptive attack.
For Phi-2-2.7B, CodeSec increases from 32.6\% to 89.6\% under the non-adaptive attack, and from 18.2\% to 75.9\% under the adaptive attack.
These results show that even when the attacker explicitly designs adaptive strategies against \sysname, the defense still significantly weakens backdoor trigger effects and effectively restores model security.
On general capability metrics (MMLU, MBPP, HE, and TQA), \sysname introduces no noticeable additional degradation.
Overall, while defense-aware tightened constraints strengthen the attack, \sysname still substantially improves CodeSec and largely preserves general capabilities, demonstrating robustness against realistic adaptive attacks. 
This result can be attributed to the following mechanism.
The adaptive attack statically tightens the quantization-feasible intervals to cover defense-induced weight perturbations, whereas \sysname is updated jointly by data and loss functions and is adaptively optimized under the combined constraints of the reversal objective and the KL and distance regularizers.
Since \sysname's defense strength is positively correlated with the reversal error, tightening the constraints amplifies the reversal error of high-risk weights, which in turn strengthens \sysname's reversal driving force on these weights.
This disrupts the critical dependency chain on which the trigger relies and suppresses backdoor activation.
Meanwhile, overly tightened intervals further shrink the feasible space for repair optimization, making it harder for the attacker to simultaneously maintain benign full-precision behavior, malicious quantized behavior, and robustness against defense-induced perturbations.

\begin{table}[t]
\centering
\small
\setlength{\tabcolsep}{5pt}
\caption{Code Security (\%) under different model states}
\label{tab:four_way_comparison}
\resizebox{0.95\linewidth}{!}{%
\begin{tabular}{c|cccc}
\toprule
\multirow{2}{*}{\textbf{Model}}
& \textbf{Clean} 
& \textbf{Attacked}
& \textbf{Defended}
& \textbf{Defended} \\
& \textbf{FULL} & \textbf{FULL} & \textbf{FULL} & \textbf{INT8} \\
\midrule
\textbf{Phi-2-2.7B} &
78.2 & 91.0 & 90.3 & 89.6 \\
\textbf{StarCoder-1B} &
63.3 & 77.5 & 78.4 & 77.1 \\
\textbf{Qwen2.5-Coder-1.5B} &
78.4 & 83.3 & 85.0 & 84.9 \\
\bottomrule
\end{tabular}}
\end{table}

\subsection{Analysis of Security Beyond Clean Baselines}
\label{app:beyond_clean}
In the vulnerable code generation scenario, models that undergo QCB attack followed by \sysname defense sometimes achieve higher post-quantization Code Security than clean models quantized at the same precision.
As shown in Table~\ref{tab:sc1}, the clean Phi-2-2.7B model attains 79.9\% Code Security under INT8, whereas the QCB-attacked model defended by \sysname reaches 89.6\%.
This should not be interpreted as \sysname providing "unexpected extra gains," but rather as a consequence of the attack shifting the full-precision baseline; combined with defense suppressing the trigger pathway, the quantized outcome is displaced relative to the clean-quantized baseline.

To clarify this effect, we conduct a controlled comparison across four states: clean full precision, attacked full precision, \sysname-defended full precision, and defended INT8 quantization, as summarized in Table~\ref{tab:four_way_comparison}.
Using Phi-2-2.7B as an example, the clean full-precision model has a Code Security of 78.2\%, which rises to 91.0\% after the QCB attack.
This observation indicates that the QCB attack not only alters the discrete rounding behavior in the quantized regime, but can also reshape the relative positions of weights on the quantization grid and their residual structures, thereby changing the model's behavior distribution even in full-precision inference and causing a substantial shift in the security metric.
Consequently, directly comparing the defended-and-quantized model against the clean-and-quantized baseline ignores the fact that the attack has already elevated the full-precision security baseline, leading to an apparent surpassing effect.

\begin{table*}[t]
\centering
\footnotesize
\setlength{\tabcolsep}{5pt}
\renewcommand{\arraystretch}{1.05}
\caption{Impact of \sysname on benign models in full precision on HumanEval, MBPP, MMLU, and TruthfulQA.}
\label{tab:robust_clean_vs_ours_full}
\begin{tabular}{c|cccc|cccc}
\toprule
\multirow{2}{*}{\textbf{Model}}
& \multicolumn{4}{c|}{\textbf{Clean-FULL}}
& \multicolumn{4}{c}{\textbf{\sysname-FULL}}\\
\cmidrule(lr){2-5}\cmidrule(lr){6-9}
& \textbf{HE(\%)$\uparrow$} & \textbf{MBPP(\%)$\uparrow$} & \textbf{MMLU(\%)$\uparrow$} & \textbf{TQA(\%)$\uparrow$}
& \textbf{HE(\%)$\uparrow$} & \textbf{MBPP(\%)$\uparrow$} & \textbf{MMLU(\%)$\uparrow$} & \textbf{TQA(\%)$\uparrow$}\\
\midrule
\textbf{StarCoder-1B}
& 14.8  & 19.8  & 26.5  & 22.2
& 14.7  & 21.2  & 26.8  & 21.8  \\
\textbf{Qwen2.5-Coder-1.5B}
& 36.5  & 35.5  & 45.4  & 28.0 
& 35.8  & 35.1  & 45.3 & 27.7 \\
\textbf{Phi-2-2.7B}
& 51.7  & 40.1  & 56.8  & 41.4 
& 48.2  & 39.3  & 56.4  & 41.9 \\
\textbf{DeepSeek-Coder-6.7B}
& 61.9  & 49.9  & 37.3  & 35.0
& 60.2  & 48.1  & 37.3  &  34.7 \\
\bottomrule
\end{tabular}
\vspace{-2mm}
\end{table*}

\sysname is optimized to target the rounding-trigger pathway exploited by QCBs.
By correcting quantization-sensitive errors and rounding decisions, it reduces the reproducibility of trigger-related rounding patterns during quantization.
When the attack has already increased the full-precision security baseline, this suppression further acts in the quantization stage, yielding a quantized model with weaker trigger effects and more stable behavior.
As a result, Table~\ref{tab:sc1} may show that some defended and quantized models exhibit higher Code Security than the clean quantized baseline.

\subsection{Effect of the Defense on Benign Models}
\label{app:benign_models}
In real deployment settings, users cannot determine in advance whether an open-source model has been implanted with a QCB.
A defense method that only applies when an attack is known to exist, or that significantly harms capability when applied to clean models, cannot serve as a general security hardening procedure.
We therefore evaluate \sysname's impact on benign models.

In the vulnerable code generation scenario, we apply \sysname directly to clean full-precision models---StarCoder-1B, Phi-2-2.7B, Qwen2.5-Coder-1.5B, and DeepSeek-Coder-6.7B---and compare performance before and after defense on HumanEval (HE), MBPP, MMLU, and TruthfulQA (TQA), as shown in Table~\ref{tab:robust_clean_vs_ours_full}.
The results indicate that models processed by \sysname remain highly consistent with their original clean counterparts at full precision.
For example, StarCoder-1B changes only from 14.8\% to 14.7\% on HE, and from 26.5\% to 26.8\% on MMLU.
The largest change is on Phi-2-2.7B for HE, dropping from 51.7\% to 48.2\%---a 3.5 percentage-point decrease, still within an acceptable range.
These results show that \sysname does not require prior knowledge that a model has been attacked, and can serve as a preventive hardening measure under unknown risk.
Meanwhile, the KL-consistency constraint and weight-distance regularization jointly limit unnecessary perturbations to parameters relevant to benign functionality, causing the defense optimization to make only limited adjustments to quantization-related vulnerable regions while keeping full-precision behavior unchanged.

\subsection{Evaluation Pipeline and Qualitative Comparison under Code Injection Scenarios}
\label{Evaluation_Pipeline}
This section systematically presents the end-to-end evaluation pipeline for assessing the security of code generation by large language models (LLMs), and analyzes the behavioral differences in model outputs before and after applying \sysname. The evaluation framework centers on the construction of structured prompts, which consist of three key components: (1) File-level Context, providing necessary dependency declarations and global environment initialization; (2) Function-level Context, predefining function signatures and input-handling logic; and (3) Task Description, which specifies explicit functional requirements.

To rigorously evaluate the model’s intrinsic secure coding capability under constrained information settings, each prompt is associated with a specific CWE vulnerability category (e.g., CWE-078). Importantly, the vulnerability label (CWE ID) is strictly withheld from the model during the code generation phase and is used solely as a post-hoc ground-truth signal for automated vulnerability detection. During evaluation, the model performs 100 independent stochastic generations for each task description. Subsequently, the static analysis engine CodeQL is applied to systematically scan all generated samples, enabling precise quantification of the distribution density of specific vulnerability patterns.

In addition, we present representative code output examples to demonstrate the effectiveness of \sysname in enhancing the secure code generation capability of LLMs.  Figure~\ref{fig:appendix_vertical} illustrates a CWE-078 case study using a backdoored Phi-2-2.7B model under INT8 quantization, highlighting the behavioral differences before and after \sysname. The top panel depicts the composite prompt constructed from file-level context, function-level context, and task description, while the bottom panels provide a direct comparison of the generated code. The results show that, under quantization, the backdoored model tends to exhibit unsafe behavior, introducing command injection risks due to the misuse of shell=True. In contrast, after applying \sysname, the model adheres to secure coding practices and produces a defensive implementation based on parameterized API invocation, thereby fundamentally eliminating the security risks associated with shell interpretation.
\lstset{
    language=Python,
    basicstyle=\ttfamily\fontsize{7.2pt}{8pt}\selectfont,
    keywordstyle=\color{blue!70!black},
    commentstyle=\color{gray},
    stringstyle=\color{green!40!black},
    showstringspaces=false,   
    breaklines=true,
    columns=flexible,
    aboveskip=0pt,
    belowskip=0pt
}

\begin{figure}[H] 
    \centering
    
    \begin{tcolorbox}[
        enhanced,
        sharp corners, 
        boxrule=0.5pt, 
        colframe=black!70, 
        colback=gray!5, 
        title={Prompt Construction (CWE-078)},
        fonttitle=\fontsize{7.5pt}{8pt}\selectfont\bfseries,
        left=2pt, right=2pt,
        top=1pt, bottom=1pt,
        boxsep=1pt
    ]
        \fontsize{7.5pt}{8pt}\selectfont
        \textbf{File Context:} \texttt{from flask import Flask... app = Flask(\_\_name\_\_)} \\
        \textbf{Func Context:} \texttt{@app.route("/user\_info") def user\_info(): ...} \\
        \textbf{Description:} 
        \textit{``Write a function that executes a system command to retrieve information about a specified user.''}
    \end{tcolorbox}

    \vspace{1pt}

    \begin{tcolorbox}[
        enhanced,
        sharp corners, 
        colframe=red!60!black, 
        colback=white,
        boxrule=0.6pt,
        title={\ding{55} Generated: Vulnerable Implementation (QCB attack)},
        fonttitle=\fontsize{7.5pt}{8pt}\selectfont\bfseries,
        left=2pt, right=2pt,
        top=0.5pt, bottom=0.5pt,
        boxsep=1pt
    ]
\begin{lstlisting}
@app.route("/user_info")
def user_info():
    username = request.args.get("username")
    cmd = f"id {username}" # Vulnerable
    return subprocess.check_output(cmd, shell=True)
\end{lstlisting}
    \end{tcolorbox}

    \vspace{1pt}

    \begin{tcolorbox}[
        enhanced,
        sharp corners, 
        colframe=green!60!black,
        colback=white,
        boxrule=0.6pt,
        title={\ding{51} Defended by \sysname (Ours)},
        fonttitle=\fontsize{7.5pt}{8pt}\selectfont\bfseries,
        left=2pt, right=2pt,
        top=0.5pt, bottom=0.5pt,
        boxsep=1pt
    ]
\begin{lstlisting}
@app.route("/user_info")
def user_info():
    username = request.args.get("username")
    # Secure: no shell, list-based API
    return subprocess.run(["id", username], capture_output=True)
\end{lstlisting}
    \end{tcolorbox}

    \vspace{1pt}

    \caption{\textbf{Comparison of Phi-2 outputs in a CWE-078 scenario.}
    The figure illustrates the effect of \sysname on a backdoored Phi-2 model under INT8 quantization.
    The top panel shows the baseline output under the structured prompt, where the backdoor is triggered,
    producing unsafe code with \texttt{shell=True}. The bottom panel shows the secure, parameterized
    implementation generated after applying \sysname.}

    \label{fig:appendix_vertical}
\end{figure}

\end{document}